\documentclass[12pt]{iopart}

\usepackage{graphicx}
\usepackage{harvard}
\usepackage{iopams}

\begin{document}
\bibliographystyle{jphysicsB}

\title[Microdosimetry of radiation field from $^{12}$C beam]{Microdosimetry of
radiation field from a therapeutic $^{12}$C beam in water: a study with Geant4 toolkit}

\author{Lucas Burigo$^{1,2,*}$, Igor~Pshenichnov$^{1,3}$, Igor~Mishustin$^{1,4}$ and Marcus~Bleicher$^{1,2}$}
\address{$^{1}$ Frankfurt Institute for Advanced Studies, Johann Wolfgang Goethe University,
60438 Frankfurt am Main, Germany}
\address{$^{2}$ Institut f\"ur Theoretische Physik, Johann Wolfgang Goethe University, 60438 Frankfurt am Main, Germany}
\address{$^{3}$ Institute for Nuclear Research, Russian Academy of Sciences, 117312 Moscow, Russia}
\address{$^{4}$ Kurchatov Institute, Russian Research Center, 123182 Moscow, Russia}

\ead{\mailto{burigo@fias.uni-frankfurt.de}, \mailto{pshenich@fias.uni-frankfurt.de}}

\begin{abstract}

We model the responses of Tissue-Equivalent Proportional Counters (TEPC) to radiation fields of
therapeutic $^{12}$C beams in a water phantom and to quasi-monoenergetic neutrons
in a PMMA phantom. Simulations are performed with the
Monte Carlo model for Heavy Ion Therapy (MCHIT) based on the Geant4 toolkit.
The shapes of the calculated lineal energy spectra agree well with measurements in both cases.
The influence of fragmentation reactions on the TEPC response to a narrow pencil-like beam with its
width smaller than the TEPC diameter is investigated by Monte Carlo modeling.
It is found that total lineal energy spectra are not very sensitive to the 
choice of the nuclear fragmentation model used.
The calculated frequency-mean lineal energy $\bar{y}_f$ differs from 
the data on the axis of a therapeutic beam by less than 10\% and 
by 10--20\% at other TEPC positions.
The validation of MCHIT with neutron beams gives us 
confidence in estimating the contributions to lineal energy spectra due to secondary neutrons produced in water 
by $^{12}$C nuclei. As found, the neutron contribution at 10~cm distance from the beam axis 
amounts to $\sim 50$\% close the entrance to the phantom and decreases to $\sim 25$\% at the depth of 
the Bragg peak and beyond it. The presented results can help in evaluating biological 
out-of-field doses in carbon-ion therapy. 
\end{abstract} 



\maketitle

\section{Introduction}

At present time the heavy-ion therapy of cancer is one of the
most advanced methods in radiation therapy~\cite{Schardt2010,Elsaesser2010,Durante2010}.
An elevated biological effectiveness of accelerated nuclei provides advantages in treatment of
radioresistant solid tumors, but also requires a thorough treatment planning to reduce undesirable
impact of radiation on healthy tissues. This should include, in particular, 
a realistic description of the production of secondary nuclear fragments. Such secondary particles
deliver dose to the tissues located farther than the Bragg peak for primary nuclei and also around
the primary beams.

Recently there were several experimental~\cite{Haettner2006} and
theoretical~\cite{Gudowska2004,Bohlen2010,Pshenichnov2010} studies which confirmed
the need to consider nuclear fragmentation reactions in heavy-ion therapy.
In particular, as shown by measurements with a 400$A$~MeV $^{12}$C beam 
(with kinetic energy of~400 MeV per nucleon), about 70\% 
of beam nuclei undergo fragmentation reactions~\cite{Haettner2006}. 
Apart from the effect of primary beam attenuation,
this means that the dose around the beam is delivered by light charged fragments,
(e.g. protons and helium nuclei) and also by neutrons. In particular,
neutrons may propagate large distances from their production points before they 
initiate secondary nuclear reactions.

As demonstrated by studies with various
transport codes, such as SHIELD-HIT~\cite{Gudowska2004}, FLUKA and
Geant4~\cite{Bohlen2010,Pshenichnov2010}, the measured attenuation of the primary carbon beam
and production of heavier fragments are generally well reproduced by the theory.
However, as follows from the same studies, there are
problems with quantitative description of the yields,
angular distributions and energy spectra for several kinds of fragments produced in
fragmentation reactions in water. In particular, the yields of alpha-particles are
underestimated. This may be related to neglecting the cluster structure of light nuclei,
which leads to the underestimation of decays of excited nuclei into channels containing alphas.
It is unlikely that such deficiency will be eliminated soon, as this will
require combining dynamical and statistical models of nuclear reactions
with sophisticated nuclear structure models into a unified, but still computationally 
effective tool.

Nuclear fragmentation models calculate the yields of secondary fragments as functions
of their charges and velocities. However, the impact of specific particle species on
living cells is rather correlated with their linear energy transfer (LET) values.
The LETs of particles with different charges $Z$, but similar velocities in first approximation are
 proportional to 
their $Z^2$. At the same time ions of different velocities and charges 
may have similar LET~\cite{Guetersloh2004}. Therefore, the reliability of nuclear 
fragmentation models, in particular used in Geant4 
calculations~\cite{Bohlen2010,Pshenichnov2010}, should be also
evaluated with respect to calculations of LET or related quantities.

The biological action of radiation is commonly characterized by microdosimetric
quantities~\cite{ICRU1983,Rossi1990,Lindborg2011} such as lineal energy $y$,
its probability density $f(y)$ and dose probability density $d(y)$.
Frequency-mean, $\bar{y}_f$, and dose-mean,
$\bar{y}_d$, lineal energies are defined as the first moments of the corresponding distributions,
see section~\ref{sec:basicsMicro} for their definitions. These distributions are routinely measured by means of  
Tissue Equivalent Proportional Counters (TEPC)~\cite{Waker1995,Waker2006}. 

In a complex radiation field different particles can eventually contribute with 
similar $y$. Therefore, the resulting $y$-distribution is build as the sum of contributions
from primary and all secondary particles passing TEPC. 
Measurements and calculations of microdosimetric quantities related to heavy-ion therapy 
were reported by several authors. In particular, microdosimetric quantities were used 
to characterize the biological dose from therapeutic $^{12}$C 
beams~\cite{Kase2006,Kase2011,Martino2010}. Similar measurements were performed 
for the first time for $^{7}$Li beams~\cite{Martino2010}.
Simulations with the PHITS code were done for a wall-less TEPC~\cite{Tsuda2010}.
The energy deposition inside and around $^{7}$Li and $^{12}$C beams in water were calculated with
SHIELD-HIT and FLUKA codes~\cite{Hultqvist2010,Taleei2011}, but without implementing
the exact geometry of the TEPC detector and, respectively, without calculating $y$-distributions. 
The FLUKA code was also used to simulate responses
of TEPC to various ions~\cite{Bohlen2011}, including $y$-distributions
for TEPCs located on the beam axis inside a water phantom irradiated by
$^{12}$C beam~\cite{Bohlen2011a}.

In our previous
papers~\cite{Pshenichnov2005,Pshenichnov2006,Pshenichnov2007,Pshenichnov2008,Pshenichnov2010,Mishustin2010}
we validated the Geant4 toolkit~\cite{Agostinelli2003,Allison2006} for use in heavy-ion therapy
simulations.  The depth-dose distributions~\cite{Pshenichnov2006} and production of
positron-emitting nuclei~\cite{Pshenichnov2007,Pshenichnov2008} were calculated by means of
the Monte Carlo model for Heavy Ion Therapy (MCHIT) based on Geant4.
Special effort was made to evaluate the quality of nuclear fragmentation
models~\cite{Pshenichnov2010,Mishustin2010} by comparing theoretical predictions with experimental data on
fragment yields~\cite{Haettner2006}.

The main purpose of the present work consists in accurate modeling of in-field and out-of-field microdosimetry
spectra, their averages and contributions from secondary neutrons for pencil-like $^{12}$C
beams, which are typical for facilities with scanned therapeutic beams. In this case a TEPC placed 
in a water phantom is impacted by a complex radiation field. It is either irradiated by a mixture of 
beam nuclei and secondary fragments or exposed exclusively to such fragments far from beam axis or 
beyond the Bragg peak.

In order to achieve the main goal of our study the following specific issues are investigated:

\begin{itemize}

\item the sensitivity of calculational results to the choice of the Geant4 models and their parameters 
for nuclear reactions and production and transport of secondary electrons, e.g. to the production thresholds;

\item the accuracy of the MCHIT model in describing microdosimetry spectra of quasi-monoenergetic neutrons
as a prerequisite for estimating the contribution of secondary neutrons from $^{12}$C beam;

\item the level of distortion of microdosimetry variables due to the impact of a focused $^{12}$C beam
with its width smaller than the TEPC diameter, i.e. when the condition of random crossing of TEPC by particles
is violated or several nuclei of fragments traverse the TEPC in a single event;

\item the dose per beam particle at various points inside the water phantom and 
the distortion of the dose field measured with TEPC due to its finite size 
and gradients of radiation field;

\item the correspondence between calculated and measured $\bar{y}_f$ and LET of beam particles for various beam 
profiles and media surrounding TEPC.

\end{itemize}

\section{Basics of microdosimetry method and relevant measurements}\label{sec:basicsMicro}

It is known that the impact of charged particles on eukaryotic cells is defined not only by
the energy imparted to the cell volume (i.e. by the absorbed dose), but also by the  
total number and spatial correlations of local energy deposits to DNA molecules inside the cell nucleus.
Such deposits lead to spatially distributed DNA lesions caused mainly by free electrons which surround
a track of a charged particle. The pattern of stochastic impacts on DNA resulting in
single, double or more complex breaks of DNA strands essentially depends on the projectile charge, mass
and velocity. It is defined by the number of electrons per unit of the track length as
well as on their energies. All these characteristics differ for photon, electron,
proton and nuclear beams. In particular, carbon
nuclei successfully used for particle therapy of cancer are
characterized by elevated biological effectiveness in the Bragg peak region as compared 
to photons and protons~\cite{Elsaesser2010,Durante2010}. Thus, these biological effects of ion 
beams should be taken into account while developing the treatment planning systems used
in particle therapy~\cite{Kase2006,Kraemer2010,Kase2011}.

Since DNA molecules are typically confined inside cell nuclei, which are structures 
of a few-micrometer size, it is desirable to study energy deposition patterns at
the micrometer scale. In particular, the in-beam microscopy is used in experiments on
irradiation of living cells. With this tool, the tracks of charged particles can be well explored, if visualized by
biological markers for DNA damage~\cite{Tobias2010}.

Since many years measurements with Tissue-Equivalent Proportional Counters
(TEPC)~\cite{ICRU1983} are considered as a practical tool for studying patterns of energy deposition 
to micrometer-size objects by various kinds of radiation.
The TEPC considered in this paper has a chamber with an inner diameter of 12.7~mm filled with low-pressure tissue-equivalent (TE) gas~\cite{Martino2010}.
It emulates a tissue-equivalent region of a micrometer size. In this way the energy 
deposited to the sensitive gas volume of this device can be related
to the energy that would be deposited to a cell nucleus~\cite{ICRU1983}. As demonstrated by 
several authors, see e.g.~\cite{Rossi1990,Lindborg2011}, microdosimetric variables are
directly related to the radiation quality factor used to quantify
risks from various kinds of radiation.

Due to the stochastic nature of particle transport in media, the amount of
energy $\epsilon$ delivered to a sensitive volume representing a cell nucleus changes from
one event to another. The value of $\epsilon$ in each 
energy deposition event can be measured by a TEPC~\cite{Waker1995,Waker2006,Lindborg2011}, as the measured electronic signal is expected to be proportional to $\epsilon$. This technique allows one
to obtain the probability distributions for lineal energy
$y=\epsilon/\bar{l}$, where $\bar{l}$ is the mean chord length of the
sensitive volume used as a detector. If one assumes that particles 
representing a homogeneous radiation field traverse a spherical detector randomly at 
various impact parameters, then $\bar{l} = 2/3 d$.

There were several experimental~\cite{Guetersloh2004,Taddei2006} 
and theoretical~\cite{Nikjoo2002,Taddei2008} studies of TEPC responses to heavy-ions. 
In these measurements and calculations the dependence of $\epsilon$ on the impact parameter of the ion
track was investigated in detail. As demonstrated by Taddei and co-authors~\cite{Taddei2008}, 
calculations based on the Geant4 toolkit can successfully reproduce the performance of a TEPC 
(Far West Technology Inc., model LET-1/2) irradiated by 
$^4$He, $^{12}$C, $^{16}$O, $^{28}$Si and $^{56}$Fe  nuclei.

Since lineal energy $y$ varies from one event to another,
the frequency-mean lineal
energy $\bar{y}_f$, which characterizes the probability density $f(y)$, can be introduced~\cite{ICRU1983}:
\begin{equation}
\bar{y}_f = \int_{0}^{\infty} yf(y){\rm d}y \ .
\label{eq:y_f_definition}
\end{equation}
The value of $\bar{y}_f$ is the first moment of $f(y)$ and serves as a measurable approximation 
for LET.

In the following also $yd(y)$-distributions, with $d(y)\equiv yf(y)/\bar{y}_f$ defined 
as the dose probability density, will be considered. The dose-mean lineal energy $\bar{y}_d$ 
is calculated as the second moment of the $f(y)$-distribution divided by its first moment:
\begin{equation}
\bar{y}_d = \int_{0}^{\infty} yd(y){\rm d}y = \frac{1}{\bar{y}_f}\int_{0}^{\infty} y^2 f(y) {\rm d}y \ .
\label{eq:y_d_definition}
\end{equation}
The value of $\bar{y}_d$ is related to the quality of radiation.

For a spherical sensitive volume filled with 
gas the total dose $D$ is given by a useful formula~\cite{ICRU1983}:
\begin{equation}
 D = \frac{0.204}{\mathit{d}^2} \bar{y}_f  \ .
\label{eq:dose_practical}  
\end{equation}
Here ${\mathit d}$ corresponds to the diameter of the sensitive volume 
expressed in $\mu$m, $\bar{y}_f$ in keV/$\mu$m and $D$ is obtained in Gy. 

Due to its compactness the TEPC device can be located at various positions inside and 
outside of phantoms irradiated by therapeutic beams.
The microdosimetry measurements with a monoenergetic 400$A$~MeV $^{12}$C beam~\cite{Endo2005} were made
together with the identification of charges of secondary fragments by scintillation counters
by the time-of-flight (TOF) method. In this way the above-described microdosimetry technique
supplements the measurements of fragment yields by particle identification made outside the phantom
with more bulky detectors, see e.g.~\cite{Haettner2006}.

Further measurements were performed with phantoms irradiated
by 290$A$~MeV $^{12}$C beam with a traditional walled TEPC~\cite{Endo2007}, also with
fragment charge identification, and later~\cite{Tsuda2010} with a wall-less counter
without tagging fragments. Recently the results of microdosimetry measurements
with 290$A$~MeV Spread Out Bragg Peak (SOBP) carbon-ion beam were also
reported~\cite{Endo2010,Kase2011}. In all these measurements the TEPCs were located at the beam axis.

Microdosimetry measurements on the beam axis as well as off-axis 
in a water phantom irradiated by a pencil-like $^{12}$C beam were performed at GSI~\cite{Martino2010}. 
Similar studies at carbon-ion and proton radiotherapy facilities with passive beam delivery
were also performed at TEPC positions outside the treatment field~\cite{Yonai2010}. 
The collected microdosimetry data challenge the theoretical methods aimed to describe particle transport in tissue-like media.

\section{Physics models used in MCHIT simulations}\label{sec:physics_models_MCHIT}

For the purposes of the present study we further developed our Monte Carlo model for Heavy-Ion Therapy
(MCHIT)~\cite{Pshenichnov2006,Pshenichnov2007,Pshenichnov2008,Pshenichnov2010,Mishustin2010}
to simulate microdosimetry measurements with TEPC devices.
The model is currently based on the Geant4 toolkit~\cite{Agostinelli2003,Allison2006}
of version 9.4 with patch 02. Input parameters for a MCHIT run are provided via a set of
user interface commands which define the beam particle, the energy distribution of the beam, its
spot size, angular divergence, as well as the dimensions and composition of the phantom
used in a particular experimental set-up.

A detailed description of the physics models included in the Geant4 toolkit is given in
the Geant4 Physics Reference Manual~\cite{G4PhysManual}. A set of models which are relevant to
a particular simulation problem should be activated by the application developer.
Such models are usually grouped in a physics list. In order to involve
certain models into simulation, one can either use the so-called predefined physics lists,
or implement customized physics lists, or even use a combination of these two options.
The predefined physics lists are provided by
Geant4 developers and distributed along with the Geant4 source code. It is convenient to use 
separate physics lists for electromagnetic and hadronic physics.

\subsection{Electromagnetic physics}

The electromagnetic processes are described by means of several predefined physics
lists of Geant4. The physics list called ``Standard Electromagnetic Physics Option 3''
is recommended by the Geant4 developers for simulations related to particle therapy.
As demonstrated~\cite{Lechner2010}, the measured positions of the
Bragg peak for carbon nuclei of various energies in water are well reproduced with 
this physics list. The involved
physics models simulate the energy loss and straggling of primary and secondary
charged particles due to interaction with atomic
electrons.  The multiple Coulomb scattering of charged particles on atomic nuclei is also
simulated. 

At each simulation step the ionization energy loss of a charged particle
is calculated according to the Bethe-Bloch formula or interpolated
between values listed in a table, depending on the particle type and energy~\cite{G4PhysManual}. 
In particular, the Bethe-Bloch formula with the shell, density and high-order corrections 
is applied in {\em G4BetheBlochModel} to protons with kinetic energy above 2~MeV.
Below 2~MeV the stopping power parameterizations~\cite{ICRU1993} are used for protons 
in {\em G4BraggModel}. The same methods are applied to alpha particles taking into
account the corresponding mass scaling with respect to proton. 
An interpolation of stopping power tables~\cite{ICRU2005} is implemented in 
{\em G4IonParametrisedLossModel} to calculate electromagnetic energy loss of ions heavier
than helium with energies relevant to our study, below 1000$A$~MeV.
The lowest kinetic energy for the production of $\delta$-electrons in this standard physic list
is 990 eV. The emission of $\delta$-electrons with energies below 990~eV is not 
simulated, but their energies are attributed to the local energy deposition.
Another predefined physics list, which extends the capability of 
electromagnetic models to produce and transport
$\delta$-electrons down to 100~eV, namely {\em G4EmPenelope} based on the Penelope 
model~\cite{G4PhysManual}, was also used for investigation of the impact of this energy threshold
for the production of secondary electrons.

In order to reduce the CPU time without affecting the accuracy of calculations, 
different cuts for production of electrons were applied in different materials. 
They are listed in table~\ref{tab:cuts}.
\begin{table}[htb]
\caption{Cut in range and energy production threshold for electrons in the phantom and
TEPC materials applied in MCHIT simulations with Standard Electromagnetic Physics (Option 3) and
with the Penelope model.}
\begin{indented}
\item[]\begin{tabular}{@{}lllll}
\br
                          & \multicolumn{2}{c}{Standard opt 3} & \multicolumn{2}{c}{Penelope}\\\ns
                          & \crule{2}                          & \crule{2} \\
 \ material               & cut in   & energy                  & cut in       & energy        \\
                          & range    & threshold               & range        & threshold     \\
                          &  (mm)    & (keV)                   & (mm)         & (keV)         \\
\mr
    water                 &   0.1    &   85.                   &   0.1        &   85.         \\
A-150 TE plastic (1.27~mm)&   0.01   &   17.6                  &   0.001      & \lineup{\0}0.3\\
 TE gas (12~kPa)          &   9      &\lineup{\0}0.99          &    3         & \lineup{\0}0.1\\         
\br
\end{tabular}
\end{indented}
\label{tab:cuts}
\end{table}

The cut in range for TE gas was defined by the lowest energy of electrons which can be transported
by the correspond models. This limitation in the electromagnetic models amounts for a cut in range of 9~mm
for the standard electromagnetic models and 3~mm for the Penelope models. Since 9~mm is of the same order as
the diameter of the sensitive volume, the comparison of results between the standard and Penelope models
helps to understand how this limitation undermines the results.

In a 1.27~mm-thin plastic shell of the TEPC detector 
the cut in range for electrons was reduced to 10~$\mu$m. This helps to simulate accurately 
grazing interactions of particles with
the plastic shell, as secondary particles produced in plastic can propagate further into 
gas. According to the default settings of Geant4 there were no cuts for 
secondary nucleons and nuclear fragments, and all produced particles (including electrons) 
after their production are transported down to zero kinetic energy.

\subsection{Modeling nuclear fragmentation}

A customized physics list was implemented for the
description of hadronic processes.  Modifications in this physics list
with respect to our previous publications concern the modeling of inelastic
nucleus-nucleus interactions. In the present study we considered two Geant4 models to describe
the first stage of nucleus-nucleus collisions, namely, the Light Ion Binary Cascade (G4BIC)
and the Quantum Molecular Dynamics (G4QMD) models. In our earlier publication~\cite{Pshenichnov2010}
the G4BIC and G4 abrasion models were employed to describe such reactions.
Recent improvements in the G4QMD code~\cite{Koi2010}, as well as the conclusions based on the
comparison of G4BIC and G4QMD~\cite{Bohlen2010}, suggest that the G4QMD model can be also
successfully used for modeling nuclear fragmentation of carbon nuclei. Therefore, in our calculations
of microdosimetric distributions both options to simulate nuclear fragmentation, G4BIC and G4QMD,
were used.

Excited nuclear fragments are frequently produced in addition to free nucleons as a result
of simulation of a nucleus-nucleus collision event by the G4BIC and G4QMD models.
Therefore, the G4ExcitationHandler of Geant4 is used to simulate subsequent decays of
excited nuclear fragments by applying various de-excitation models depending on
the mass and excitation energy of these fragments.

The Fermi break-up model (G4FermiBreakUp) is applied to
nuclei lighter than fluorine. It is designed to describe explosive disintegration of
excited light nuclei~\cite{Bondorf1995} and it is highly relevant to collisions of light
nuclei with nuclei of tissue-like materials. As demonstrated~\cite{Mishustin2010}
G4BIC linked with G4FermiBreakUp better describes the production of
lithium and beryllium nuclei, as well as secondary neutrons by $^{12}$C nuclei in water and PMMA
compared to the option of G4BIC linked with the nuclear evaporation model. As demonstrated by B\"ohlen et al.~\cite{Bohlen2010} the combination of G4QMD and G4FermiBreakUp also describes data well. Therefore, we used
G4FermiBreakUp to de-excite light fragments also in the present work.

\subsection{Hits and deposited energy}

We use the primitive scorer classes of Geant4 to calculate physical quantities which
characterize the TEPC response in each event. 
The energy $\epsilon$ deposited to the TEPC sensitive volume is calculated by means of the {\em G4PSEnergyDeposit} scorer.  
This scorer stores a sum of particles' energy deposits to the sensitive volume in each event. 
The number of tracks $N$ that pass through the TEPC sensitive volume, but do not start or stop inside it, is calculated by employing {\em G4PSPassageCellCurrent}.
In particular, this means that numerous low-energy $\delta$-electrons which are produced 
by energetic nuclei and stopped inside the TE gas are not counted by this scorer.
Finally, {\em G4PSPassageTrackLength} is used to calculate the total track length $l$ inside the 
TEPC sensitive volume in single and multi-particle events. With this scorer 
the track length is defined as the sum of step lengths of the particles inside the 
volume, and again only tracks which traverse the volume are taken into account. 
This means that newly-generated or stopped tracks inside the volume
are excluded from the calculation of $l$. The employed {\em G4PSPassageTrackLength} scorer is thus
insensitive to those secondary electrons which were produced by a fast particle inside the 
TEPC sensitive volume, but stopped inside it.
The total track length is scored in order to evaluate the mean track length through the
detector for comparison with the mean chord length $\bar{l} = 2/3 d$ in the standard
conditions for microdosimetry measurements.

\section{Modeling experimental set-up}\label{sec:setup}

In this work we calculate microdosimetric distributions and compare them with results of
two experiments~\cite{Nakane2001,Martino2010}. In the first experiment a
PMMA phantom was irradiated by broad quasi-monoenergetic neutron beams 
at the Japan Atomic Energy Research Institute (JAERI). In the second experiment a water phantom was
irradiated by carbon and lithium pencil-like beams at GSI. The same type of
TEPC device (Far West Technology Inc., model LET-1/2) was
used by both groups. This allowed us to validate our calculational approach for
this specific TEPC model first with neutron irradiation data~\cite{Nakane2001}, and then extend it
to the GSI measurements~\cite{Martino2010} with $^{12}$C beam.
In a previous study~\cite{Mishustin2010} we demonstrated that MCHIT can estimate reasonable well the
energy and angular spectra of secondary neutrons produced by 200$A$~MeV carbon beam in thick water target.
Therefore, if MCHIT is able to reproduce microdosimetry spectra for neutrons measured with a TEPC
we believe that the contribution of secondary neutrons to the microdosimetry spectra for carbon beam
measured with the same TEPC device can be estimated with confidence.

The actual TEPC dimensions and the chemical composition of TEPC materials were implemented 
as accurate as possible in our simulations. It was assumed that the tissue equivalent gas was propane
C$_3$H$_8$ (55\%) with addition of CO$_2$ (39.6\%) and N$_2$ (5.4\%). A gas container
was defined in simulations as a shell of tissue-equivalent plastic composed of H (10.1\%), 
C (77.6\%), N (3.5\%), O (5.2\%), F (1.7\%) and Ca (1.9\%), in accordance with the  
properties of the NIST material Shonka A-150 plastic. An aluminum outer shell of the 
TEPC was also introduced, although simulation results were found to be rather insensitive to
the presence of the Al shell because of its small thickness. The central wire located 
inside the gas sphere was not considered in the simulations. Following Taddei et al.~\cite{Taddei2006},
it was assumed that quite rare events in which a projectile particle interacted with this wire 
are rejected in measurements. We also neglect the bending of the simulated tracks due to the voltage applied 
to the TEPC.

Experimental results for a TEPC located
at 5~cm depth inside the PMMA phantom irradiated by 40~MeV and 65~MeV quasi-monoenergetic
neutrons were selected from JAERI data to validate the MCHIT model.
The pressure of the tissue equivalent gas was 9.03~kPa which emulated a tissue 
sphere of 2.07 $\mu$m in diameter. 

In the second experiment~\cite{Martino2010} a water phantom
of $30\times 30\times 30$~cm$^3$ (including side walls made of PMMA) was
irradiated by a 300$A$~MeV $^{12}$C beam.
In the simulations the parameters of the $^{12}$C beam used at GSI were implemented as accurate as possible. 
The beam had a concentric spot with a size of 3~mm FWHM at beam exit window, an angular distribution 
with the Gaussian profile of 1~mrad FWHM, and an energy distribution with the Gaussian profile of 0.2\% FWHM.
The $yd(y)$-distributions per beam particle measured at nine points
inside the water phantom were reported~\cite{Martino2010}. They correspond to
0~cm, 2~cm and 10~cm radii and to three values of depth: at the plateau of the depth-dose curve,
at the Bragg peak and in the tail region. In this experiment water was enclosed in a container with a 20~mm thick PMMA wall. Carbon nuclei also penetrated through a vacuum window made of aluminum, a scintillation detector
and a parallel-plate ion chamber installed in front of the phantom. 
These beam-line elements and PMMA walls of the phantom were estimated by Martino et~al. 
as equivalent water thickness of 25.1~mm. This thickness was applied to the TEPC depth in order to compare the data 
with simulation results, since calculations were performed with an equivalent wall-less
water phantom of $30\times 30\times 30$~cm$^3$. 
It was found that a 2~mm shift to a deeper position of the TEPC is required in order to reproduce
the peak in the $yd(y)$ spectrum for the TEPC positioned in the vicinity of the Bragg peak
on the beam axis as well as the depth-dose profile. At this position the microdosimetry spectrum
is highly sensitive to a few millimeters shift of the device. One can attribute this shift to  
uncertainties in the measurements of TEPC positions, beam energy and 
the equivalent water thickness in front of the water phantom.
Some inaccuracy of the electromagnetic physics models of Geant4 can not be excluded as well. 
For consistency this shift was 
applied to all TEPC positions as a systematic correction.
The gas pressure was set to the experimental
value of 12~kPa which corresponds to a sphere of tissue of 2.75~$\mu$m in diameter.
The TEPC positions used in our simulations
are listed in table~\ref{tab:TEPCpositions}.

Using the axial symmetry of the set-up in the second experiment we replaced a single physical TEPC by a ring of
identical virtual TEPCs located at the same depth and radial distance from the beam axis.
During each run the lineal energy events were scored in a histogram
corresponding to a particular ring.  At the end of the run
the number of events in each lineal energy bin were divided by the number of virtual counters in the ring and thus became equivalent to a single physical counter.
With this method one can increase the simulated number of hits 
for those positions of counters where hits are scarce.  
This is particularly important for TEPCs located at large distances, e.g., at 10~cm, 
from the beam axis. The number of counters in each ring and the positions of the rings with
respect to the beam axis were chosen to avoid any crosstalk impacts on the virtual detectors. 
This was done to satisfy the condition that the particles produced
inside one of the virtual counters, or which have just traversed it, should not hit other counters.
Therefore, calculations for the nine TEPC positions listed in table~\ref{tab:TEPCpositions}
were split into three independent runs.
During the run ``I'' three outer rings of TEPCs of 10~cm in radius
each consisting of 24 counters were placed in the water phantom. In the same run a counter on the
beam axis and a ring of 2~cm in radius consisting of four TEPCs were placed at 27.71~cm depth. 
Each of the runs ``II'' and ``III'' was executed with a ring
of 2~cm in radius consisting of  four counters and with a single counter located 
at the same depth in the center of the ring. Typically, the histories of 
$10^7$ to $5\times 10^7$ $^{12}$C nuclei traversing the phantom were simulated in each run.

\begin{table}[htb]
    \caption{Positions of the TEPC counters inside the water phantom
used in MCHIT simulations and their labeling in the following text and figures.
TEPCs were grouped into rings with a number of counters in each ring depending on the ring radius.
Each TEPC position was modeled in a certain run labeled in the last column.}
\begin{indented}
\item[]\begin{tabular}{@{}lllll}
\br
   radius        & depth         & TEPC's position        & number of  &  run \\
   (mm)          & (mm)          &  notation              & counters   & label\\
                 &               &                        & in the ring&      \\
\mr
  \lineup{\0\0}0.&\lineup{\0}52.1&\lineup{\0}0 cm, plateau&\lineup{\0}1& III  \\
  \lineup{\0\0}0.&          179.1&\lineup{\0}0 cm, peak   &\lineup{\0}1& II   \\
  \lineup{\0\0}0.&          277.1&\lineup{\0}0 cm, tail   &\lineup{\0}1& I    \\
   \lineup{\0}20.&\lineup{\0}52.1&\lineup{\0}2 cm, plateau&\lineup{\0}4& III  \\
   \lineup{\0}20.&          179.1&\lineup{\0}2 cm, peak   &\lineup{\0}4& II   \\
   \lineup{\0}20.&          277.1&\lineup{\0}2 cm, tail   &\lineup{\0}4& I    \\
             100.&\lineup{\0}52.1&          10 cm, plateau&          24& I    \\
             100.&          179.1&          10 cm, peak   &          24& I    \\
             100.&          277.1&          10 cm, tail   &          24& I    \\

\br
\end{tabular}
\end{indented}
\label{tab:TEPCpositions}
\end{table}

\section{TEPC response to quasi-monoenergetic neutron beams}\label{sec:neutron_measurements}

It is expected that radiation fields far from the beam axis
have a large contribution from neutrons produced in fragmentation of beam nuclei.
Therefore, before performing microdosimetry simulations with nuclear beams we checked the validity of
the MCHIT model with respect to irradiation of the same TEPC model by neutrons. 
The data obtained with quasi-monoenergetic neutron beams~\cite{Nakane2001}
were used for this purpose. Energy spectra of neutrons used in the two sets of
measurements~\cite{Nakane2001} were modeled as superpositions of Gaussian peaks centered
at 40 and 65~MeV, respectively, and broad plateaus. The ratios between the flux of neutrons
with energies within the peak and the total neutron flux were set to 1:2.768 and 1:2.807
for 40 and 65 MeV spectra, respectively, following the estimations by Nakane et al.~\cite{Nakane2001}.
As proved by our simulations, the calculated $yd(y)$ distributions are not very
sensitive to these ratios. For example, the calculations with these ratios set to 1:3.5
provide results similar to ones obtained with original ratios~\cite{Nakane2001}.  
In the following the above-described spectra
composed of peaks and plateaus are referred to as ``40~MeV'' and ``65~MeV'' neutrons.

\begin{figure}[htb]
\begin{centering}
\includegraphics[width=1.0\columnwidth]{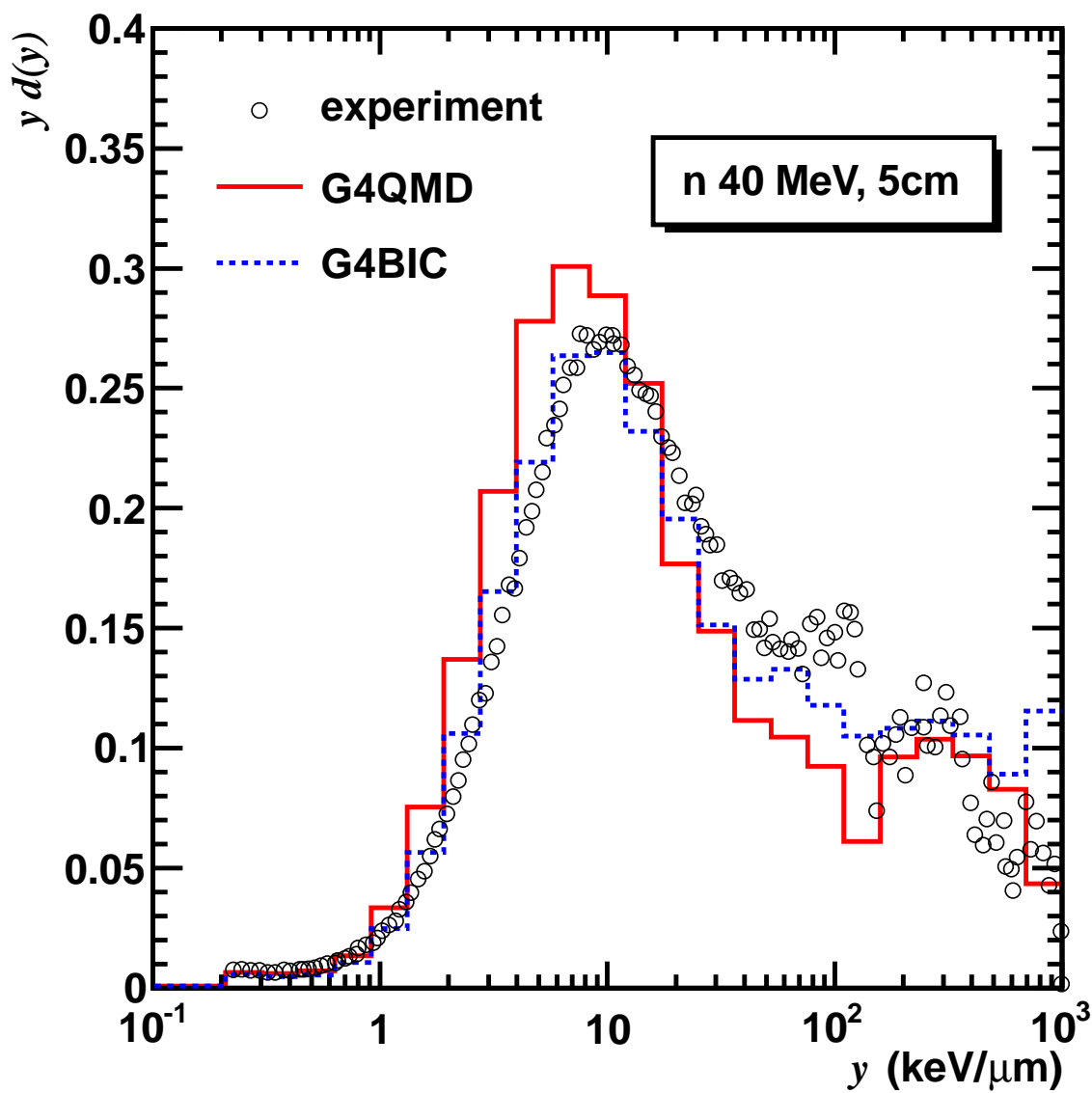}
\end{centering}
\caption{Calculated microdosimetric spectra for a TEPC at 5~cm depth inside
a PMMA phantom irradiated by quasi-monoenergetic ``40 MeV'' neutrons.
Theoretical results obtained with
G4QMD and G4BIC models are presented by solid and dashed histograms, respectively.
Experimental data~\protect\cite{Nakane2001} are shown by points.
}
\label{fig:40MeV_n}
\end{figure}

MCHIT results for a TEPC located at 5~cm depth inside a
PMMA phantom of $30\times 30\times 30$~cm$^3$ are shown in figure~\ref{fig:40MeV_n}
together with the experimental data~\cite{Nakane2001}.
In this experiment the phantom was irradiated by  ``40~MeV'' neutrons,  and the TEPC
was located on the axis of the irradiation field.
Neutrons were delivered to the phantom via a collimator of 10.9~cm in diameter, which was
much larger than the diameter of the TEPC. Since neutrons traversed the TEPC randomly 
at various impact parameters, the mean neutron path $\bar{l}$ inside the TEPC
sensitive volume amounted to $2/3$ of its diameter. As explained in section~\ref{sec:basicsMicro},
this corresponds to the standard conditions of microdosimetry measurements~\cite{ICRU1983}. 

\begin{figure}[htb]
\begin{centering}
\includegraphics[width=1.0\columnwidth]{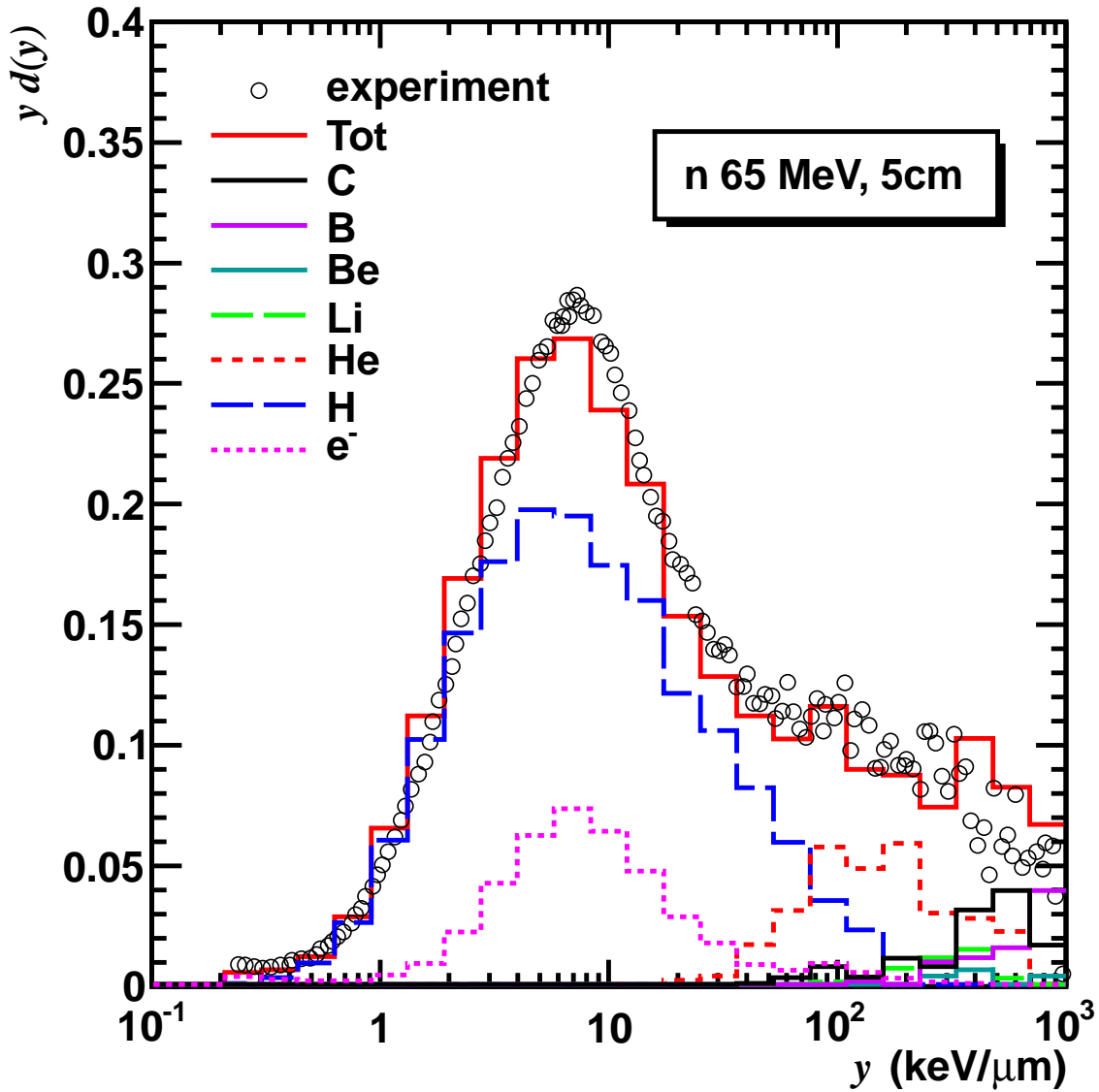}
\end{centering}
\caption{Calculated microdosimetric spectra for a TEPC at 5~cm depth inside
a PMMA phantom irradiated by quasi-monoenergetic ``65 MeV'' neutrons.
Results of MCHIT calculations with G4BIC model are shown.
In addition to the total $yd(y)$ distribution the contributions of
specific particles are shown by various histograms as explained on the legend.
Experimental data~\protect\cite{Nakane2001} are shown by points.
}
\label{fig:65MeV_n_cont_BIC}
\end{figure}

\begin{figure}[htb]
\begin{centering}
\includegraphics[width=1.0\columnwidth]{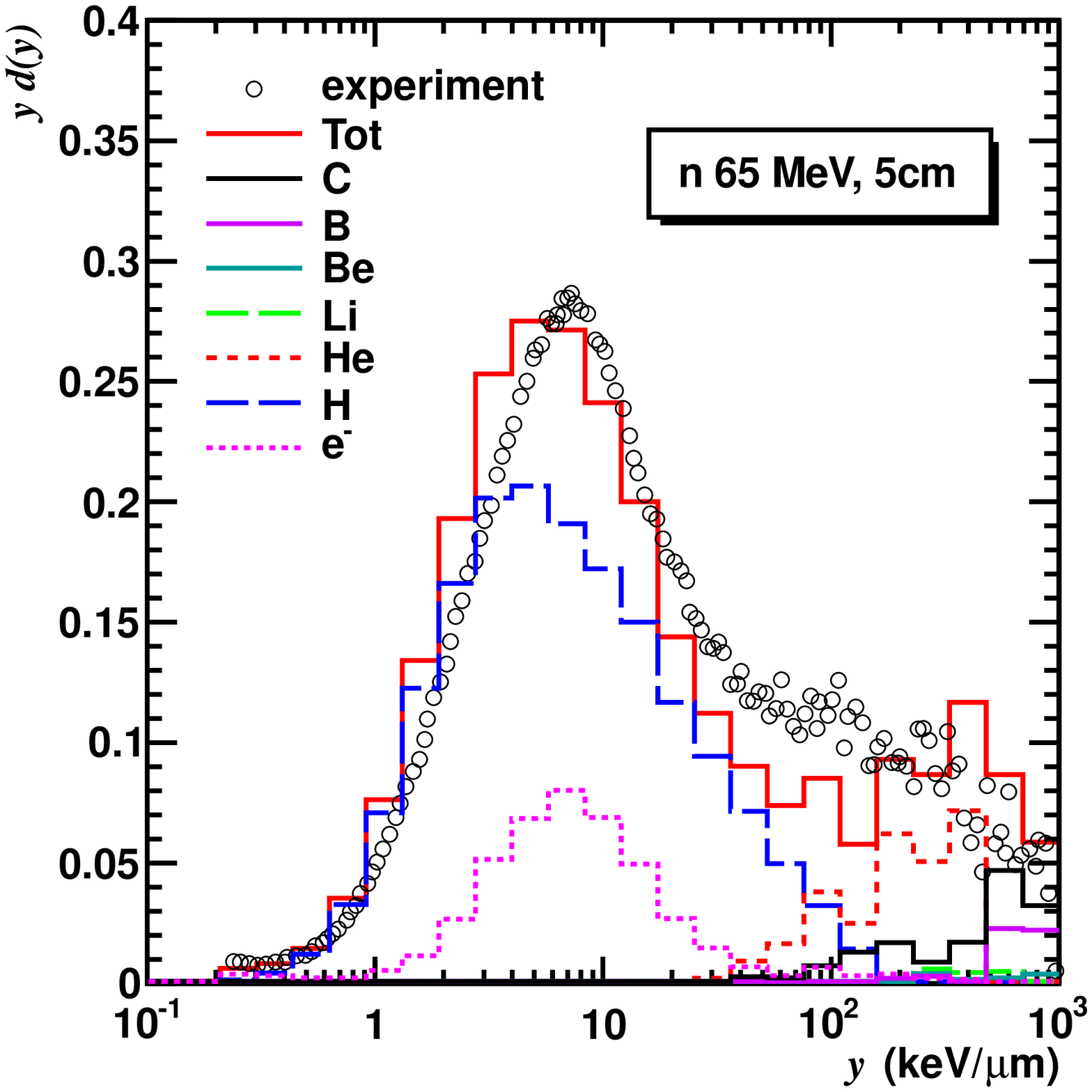}
\end{centering}
\caption{Same as in figure~\ref{fig:65MeV_n_cont_BIC}, but for MCHIT results with G4QMD model.
}
\label{fig:65MeV_n_cont_QMD}
\end{figure}

The results presented in figure~\ref{fig:40MeV_n} were obtained by simulating $4~\times~10^8$ 
primary neutrons for each of the runs where either G4QMD or G4BIC were used 
to simulate neutron-induced nuclear reactions. The number of primary particles was such that the
fluctuations in the histogram is of the same order as those presented by the experimental data. No
error bars are presented due to the lack of error bars in the reported experimental values.
The microdosimetry parameters computed from the spectra are $\bar{y}_f = 7.54$~keV/$\mu$m 
($\bar{y}_d = 84.2$~keV/$\mu$m) for the experiment and  
$7.55$~keV/$\mu$m ($110$~keV/$\mu$m) and $6.21$~keV/$\mu$m ($80.2$~keV/$\mu$m) for the 
simulations with G4BIC and G4QMD, respectively.
Microdosimetry spectra calculated for 
the same set-up,
but for the irradiation with ``65~MeV'' neutrons are shown separately in figures~\ref{fig:65MeV_n_cont_BIC}
and~\ref{fig:65MeV_n_cont_QMD}.
The corresponding microdosimetry parameters are  $\bar{y}_f = 5.51$~keV/$\mu$m 
($\bar{y}_d = 75.2$~keV/$\mu$m) for the experiment and $5.33$~keV/$\mu$m ($86.1$~keV/$\mu$m) 
and $4.75$~keV/$\mu$m ($84.2$~keV/$\mu$m) for the simulations with G4BIC and G4QMD, respectively.

Neutrons do not transfer their energy directly to the TEPC gas volume, but rather through 
secondary charged particles produced in neutron interactions with nuclei in the gas cavity, TEPC 
wall or even in surrounding layers of PMMA.  
Several recoil particles can be produced by a single neutron. MCHIT makes it possible to
estimate partial contributions from each kind of recoil particles to the total $yd(y)$ curve.
Such contributions were calculated as following.
Firstly, the amount of energy deposited to the sensitive volume by every kind of particle is 
scored separately at each lineal energy event. Secondly, the relative contribution 
to $yd(y)$ for each kind of particle in this event is calculated from the ratio 
between the energy deposited by particles of a given kind and the total energy 
deposited in the event. The partial contributions from specific particles to  $yd(y)$
for the whole run are shown in figures~\ref{fig:65MeV_n_cont_BIC}
and~\ref{fig:65MeV_n_cont_QMD}.

Contributions of specific secondary particles to the $yd(y)$ distribution for ``65~MeV''
neutrons are shown in figure~\ref{fig:65MeV_n_cont_BIC}. 
These distributions were obtained with G4BIC applied to simulate neutron-induced nuclear reactions.
This figure demonstrates the origins of a peak
at $y\sim7$~keV/$\mu$m and a shoulder extending to higher $y$ values. The peak is due
to recoil protons, while the shoulder is formed by more heavy recoil nuclei,
mostly by alpha-particles ($y>30$~keV/$\mu$m) and carbon nuclei ($y>100$~keV/$\mu$m).
The contributions from boron, lithium and beryllium nuclei are much smaller and attributed mostly to
$y>200$~keV/$\mu$m. 
The microdosimetric distributions for ``40~MeV'' neutrons (not shown) 
were calculated in a similar way. In this case the maximum is located at 
$y\sim9$~keV/$\mu$m due to less energetic primary neutrons.
Therefore, neutrons transfer less energy to recoil protons, which get slower and therefore have higher LET values.

Contributions of various recoil particles to $yd(y)$ distribution calculated
with G4QMD for ``65 MeV'' neutrons are shown in figure~\ref{fig:65MeV_n_cont_QMD}.
As seen from the figure, MCHIT predicts more fast recoil protons (with lower LET) compared
to experimental data. This leads to a shift of the peak of the calculated distribution 
to lower $y$. Regarding recoil nuclei, 
results obtained with G4QMD model differ from measurements at
$y\sim$100~keV/$\mu$m where energy deposition events are mostly caused by helium nuclei.
This can be explained by an underestimation of alpha-particle yields in neutron-induced
reactions simulated by this model.

As shown in this section, the MCHIT model with G4BIC applied to simulate neutron-induced reactions
demonstrates a better agreement with experimental data compared to the MCHIT with G4QMD.
The maximum of the theoretical $yd(y)$ distributions calculated with G4QMD is slightly
shifted to lower $y$ values with respect to the measurements~\cite{Nakane2001}.
The proton edge at $y\sim100$~keV/$\mu$m is better reproduced by G4BIC.
The measured $\bar{y}_f$ is well
reproduced by G4BIC, but $\bar{y}_d$ is overestimated for ``40 MeV'' neutrons 
due to a higher probability of large size events. G4QMD underestimates the frequency-mean 
lineal energy by 14--18\% 
but better reproduces the dose-mean lineal energy compared to G4BIC.
Nonetheless, the overall shape of the microdosimetry spectra and 
the magnitude of microdosimetry parameters for 
quasi-monoenergetic neutrons are reasonably well reproduced by MCHIT using both  
G4BIC and G4QMD models.
One can note a better agreement between
the measured~\cite{Nakane2001} $yd(y)$ distributions and MCHIT calculations 
as compared with the PHITS code
used for modeling~\cite{Tsuda2007} the same TEPC measurements~\cite{Nakane2001}.
This can be explained by the fact that in those PHITS
calculations~\cite{Tsuda2007} the production and transport of $\delta$-electrons
was neglected, while these processes are taken into account in our MCHIT 
simulations.

\section{TEPC response to a therapeutic $^{12}$C beam}\label{sec:response_to_c12}

Now we turn to the central part of our study where we use MCHIT to model the response of the TEPC to
a pencil-like therapeutic $^{12}$C beam. We start with calculating the numbers of particles which traverse the
TEPC placed at various positions in a water phantom and then continue with the distributions of the total track
length inside the TEPC and with microdosimetry spectra.

\subsection{Number of particles which cross the TEPC}\label{sec:topology}

As described above in section~\ref{sec:basicsMicro}, the lineal
energy of a particle traversing a spherical TEPC can be estimated  
by dividing the energy $\epsilon$ imparted to the detector by the mean chord length 
$\bar{l}=2/3 d$, where $d$ is the TEPC diameter. This estimation is based on the assumption that the 
TEPC is placed in a homogeneous radiation field and traversed by a single particle per event.
This corresponds to the simplest standard event topology characterized by a single track 
with $l\leq d$. This means that the particle traverses the TEPC sensitive volume by a 
straight-line trajectory, it is not stopped inside the volume, and secondary particles are 
not produced neither in the volume nor in the vicinity of it.

In the experiment~\cite{Martino2010} the TEPC was irradiated either directly by a focused
$^{12}$C beam, or hit only by secondary particles at TEPC locations outside the beam spot.
In this section we investigate the event topology relevant to the both cases.
Indeed, on the contrary to the measurements with neutron beams described in
section~\ref{sec:neutron_measurements}, the size of the $^{12}$C beam was essentially
smaller (3~mm FWHM) compared to the diameter of the TEPC sensitive volume ($d=12.7$~mm).
Therefore, for a TEPC placed on the beam axis at the plateau region of the depth-dose 
distribution and, possibly, close to the Bragg peak, the mean chord length will 
be larger than $2/3d$ due to mostly central beam incidence.
Moreover, several secondary particles can impact the TEPC in a single event.
In this section we investigate whether the irradiation conditions match
the standard ones for which the mean chord length $2/3d$ was used 
to calculate lineal energy $y$ for a single particle traversal.

\begin{figure}[htb]
\begin{centering}
\includegraphics[width=1.0\columnwidth]{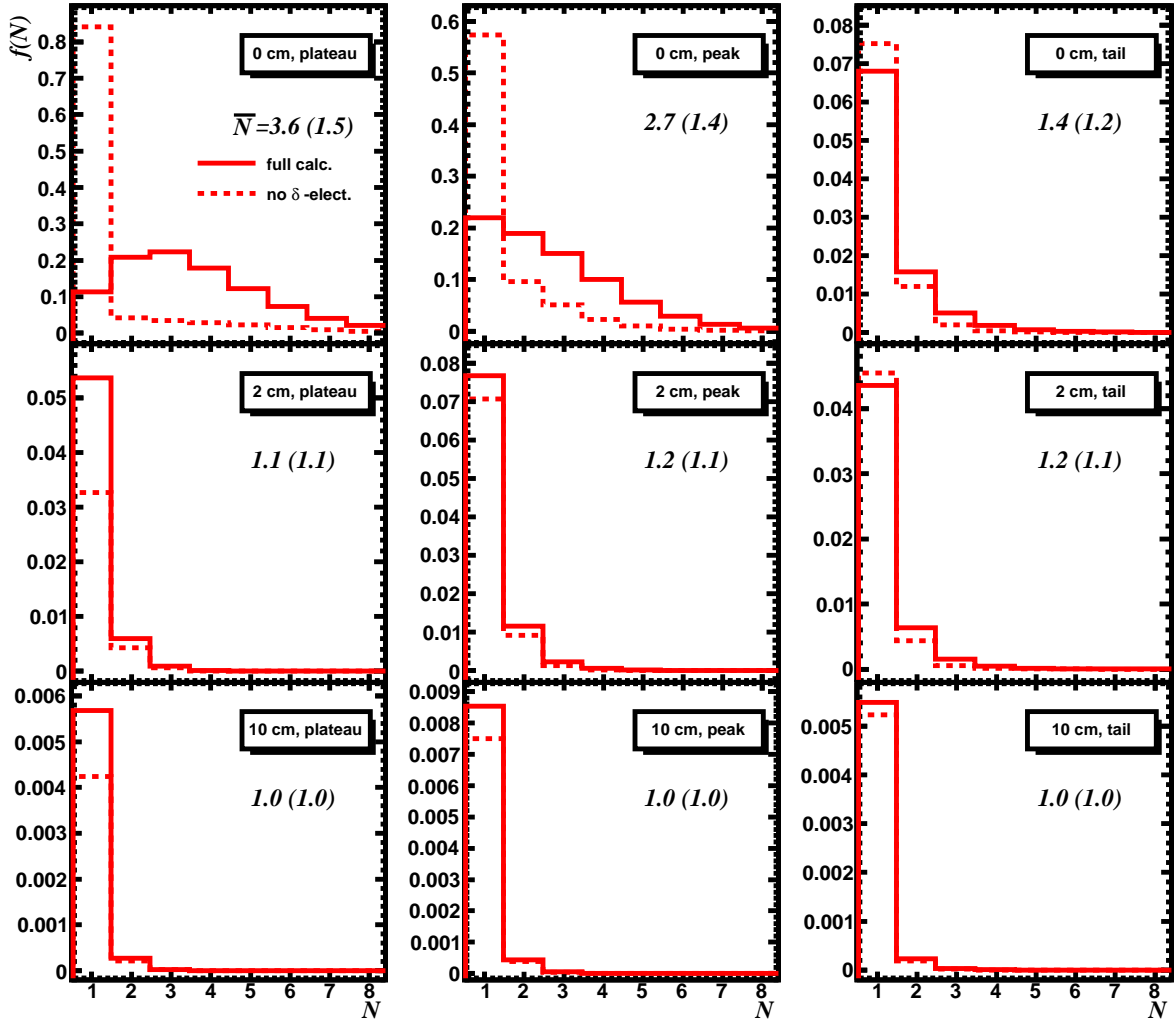}
\end{centering}
\caption{Probability distributions for the number of tracks 
in TEPCs at different locations in the water phantom irradiated by 
300$A$~MeV $^{12}$C beam. The condition to have at least one track in the corresponding TEPC 
is applied. The distributions with and without simulation of 
$\delta$-electrons are presented by solid- and dashed-line histograms, respectively, and 
calculated per beam particle. The average number of tracks $\bar{N}$ per 
traversing event are presented for each TEPC location, and the values obtained 
without $\delta$-electrons are given in parentheses.}
\label{fig:number_of_tracks}
\end{figure}

Since a TEPC detector is only sensitive to the total energy $\epsilon$ deposited to
its volume given by the number of ionized atom-electron pairs produced in the gas chamber, 
the number of tracks in each event and their length (event topology) can not 
be identified in experiment. However, one can retrieve this information from MCHIT simulations 
by scoring the number of tracks and their total length inside the TEPCs. 
The probability distributions for the number of tracks $f(N)$ inside the TEPCs are shown in 
figure~\ref{fig:number_of_tracks}. The distributions are calculated for the events with at least one 
particle traversing the TEPC, and they are given at nine positions in the water phantom 
listed in table~\ref{tab:TEPCpositions}.
These distributions are calculated per single beam particle 
and represent the probability for a TEPC placed at a certain position in the phantom to 
be traversed by $N$ particles of any kind, where $N=1$ corresponds to the standard event topology defined above
despite of multiple $\delta$-electrons stopped inside the TEPC. 
The tracks of the particles which do not deposit energy to the TEPC are also included.  

The distributions presented in figure~\ref{fig:number_of_tracks} were calculated with and without
production of $\delta$-electrons using the standard electromagnetic models.
In the latter case only secondary nucleons and nuclear fragments 
were produced and counted while they traversed TEPCs. As one can see from the distribution calculated at
``0~cm, plateau'', the TEPC at this location is traversed by beam particles with 
$\sim 0.85$ probability. This reflects the attenuation of the primary beam in the first
5~cm of water. On the contrary, the probability to hit the TEPC at ``10~cm, tail'' position far 
from the beam axis is only $\sim 0.0053$ per beam particle.
As deduced from the comparison of the distributions calculated with and without 
generating $\delta$-electrons,  all TEPCs placed at the beam axis are usually 
traversed by several electrons in addition to nucleons and nuclei. 

The topology of each event associated with the traversing of TEPC by one or several particles, 
can be characterized by the average number of tracks $\bar{N}$ calculated per event. 
These average numbers are presented in each panel of figure~\ref{fig:number_of_tracks} for both 
calculational options -- with and without $\delta$-electrons.
When the production of $\delta$-electrons is neglected, the most probable events 
are characterized by a single particle track, with $\bar{N}=1.5$ and $1.4$ under the direct 
impact of the beam at the plateau and peak, respectively. 
However, when the production of $\delta$-electrons is considered, $\bar{N}=3.6$ at the plateau and
$\bar{N}=2.7$ at the Bragg peak position. Although these electrons deliver less energy to the 
TEPC when compared to beam nuclei, their contribution can affect the results of simulations.  
As demonstrated in figure~\ref{fig:number_of_tracks}, the $N$-distributions and $\bar{N}$ 
calculated for other seven positions are less affected by neglecting $\delta$-electrons.
At these positions the TEPCs are not directly impacted by the $^{12}$C beam.
When a lower energy threshold of 100~eV is used with the Penelope models, the average number of tracks
on the beam axis is increased at the plateau to 3.7 and the Bragg peak position to 3.1, but for all
other positions $\bar{N}$ does not change.
  
\begin{figure}[htb]
\begin{centering}
\includegraphics[width=1.0\columnwidth]{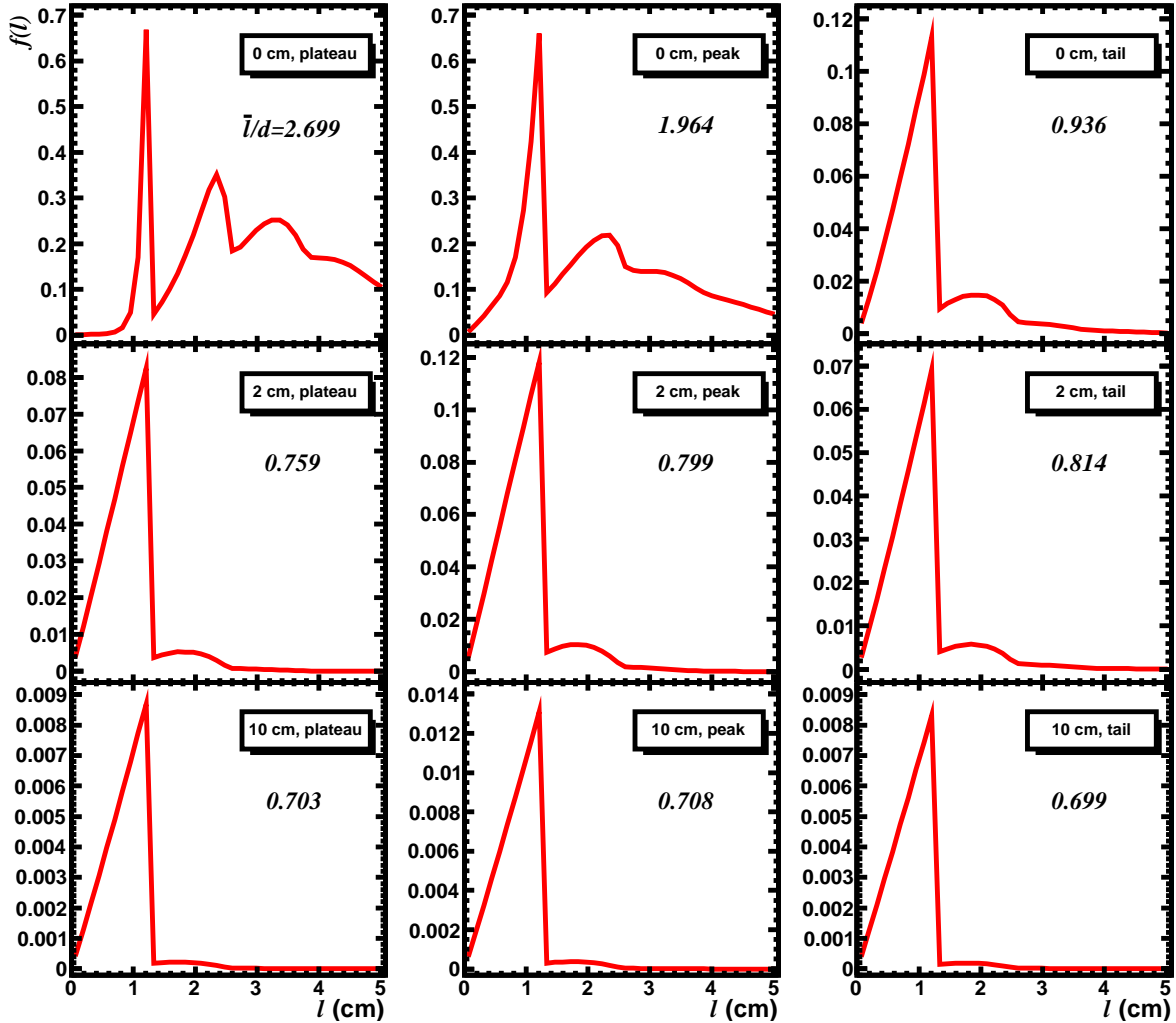}
\end{centering}
\caption{Probability distributions for the total track length for particles traversing 
the TEPCs at different locations in the water phantom irradiated by 
300$A$~MeV $^{12}$C beam. The condition to have at least one track in the TEPC is
applied. The calculations were performed with production of $\delta$-electrons using
the standard electromagnetic models.
The average values of $\bar{l}/d$ calculated per event are listed for each TEPC position.}
\label{fig:track_length}
\end{figure}

The probability distributions $f(l)$ for the total track length $l$ 
calculated with {\em G4PSPassageTrackLength} 
for particles traversing the TEPC at different locations in the water 
phantom are shown in figure \ref{fig:track_length}.
The values of $\bar{l}/d$ calculated per event of TEPC traversing are also given for each 
TEPC position. The irradiation of TEPCs by a thin $^{12}$C beam is clearly reflected
in the $f(l)$-distributions calculated for ``0~cm, plateau'' and ``0~cm, peak''. 
As one can note, these distributions present a pronounced peak at the track length corresponding
to the diameter of the gas cavity and a broad tail for higher track lengths 
with $\bar{l}/d$ much larger than $2/3d$ due to secondary electrons and beam fragments.
These distributions contrast to the linear dependence 
of $f(l)$ versus $l$ (for $l<d$) found at all other TEPC positions which indicates a random particle
incidence in the cavity. 
At ``0~cm, tail'' $\bar{l}/d$ is still large, $\bar{l}/d \sim 0.936$, due to multiple tracks in a single event.
Some events with two and three nucleons or nuclei in the TEPC are characterized by $l>d$ at
``0~cm'' positions, as already demonstrated earlier in figure~\ref{fig:number_of_tracks}.  
The distributions at the radius of 2~cm still 
exhibit the presence of one and two tracks of nucleons or nuclei in the TEPC.
The mean chord length is close to $2/3d$ only for the TEPCs 
at the radius 10~cm. There the probability of an event with 
two particles in the TEPC is negligible.

The analysis of $f(N)$ and $f(l)$ distributions given in this section demonstrates different
topologies of events in the TEPCs placed at the beam axis and far from the beam.
In the latter case the events are mostly characterized by the simplest standard topology 
associated with a single track of nucleon or nuclear fragment traversing the TEPC sensitive volume. 
On the contrary, the events in the TEPCs placed on the beam axis
at the entrance to the phantom (``0~cm, plateau'') and at the Bragg peak (``0~cm, peak'') are
more complicated.  A part of events is characterized by one or two tracks of beam 
nuclei, secondary nucleons and/or  nuclear fragments  traversing the TEPC 
sensitive volume. In addition, tracks of two, three or more energetic electrons are 
frequently present in the TEPCs placed on the beam axis.

\subsection{Lineal energy spectra inside the water phantom}\label{sec:spectra}

\begin{figure}[htb]
\begin{centering}
\includegraphics[width=1.0\columnwidth]{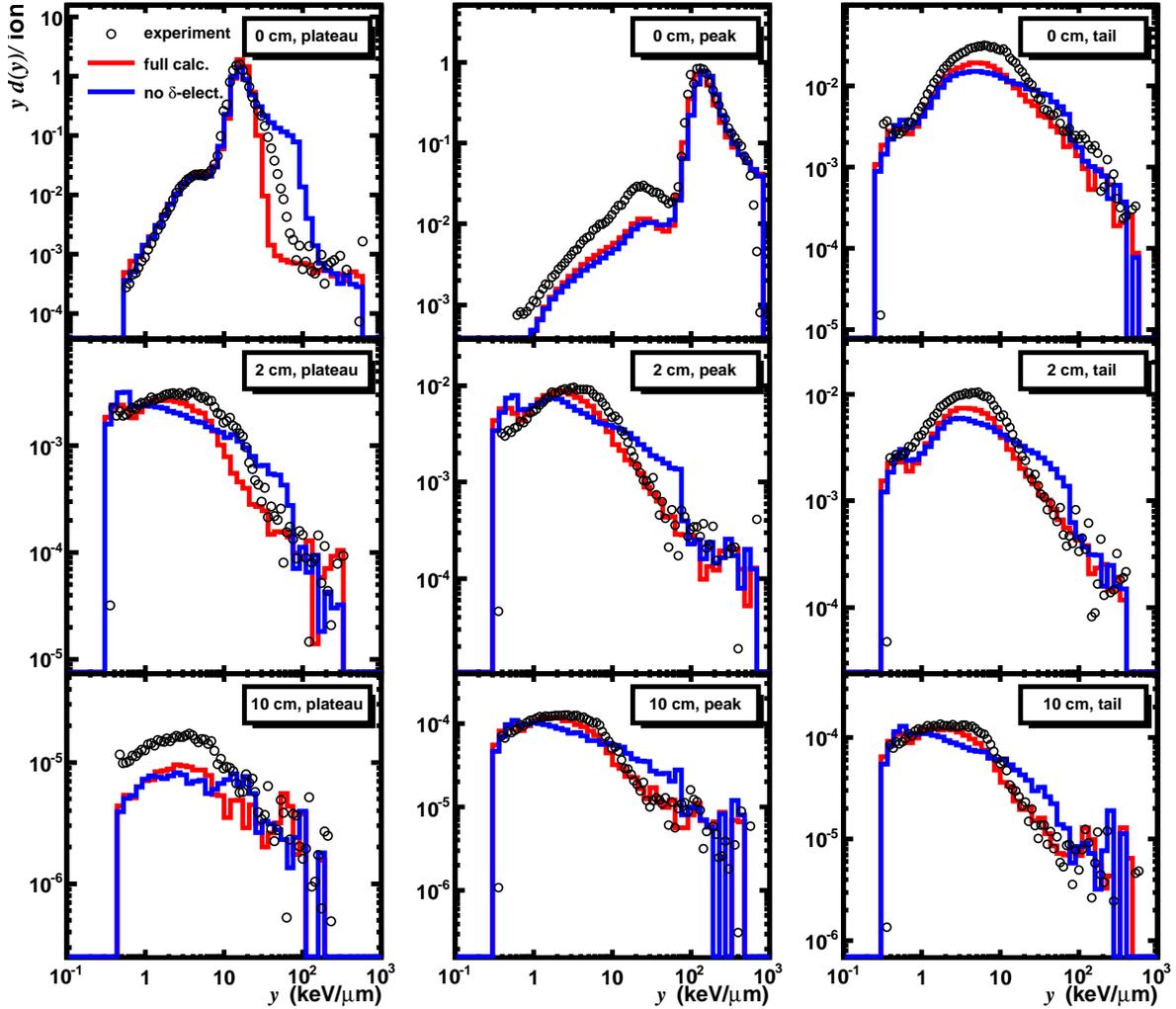}
\end{centering}
\caption{Microdosimetric $yd(y)$ distributions in water phantom irradiated by
 300$A$~MeV $^{12}$C nuclei. Distributions calculated by MCHIT per beam particle 
with and without simulating $\delta$-electrons are presented by red- and 
blue-line histograms, respectively. Points represent experimental data~\cite{Martino2010}.}
\label{fig:with_wo_delta_el}
\end{figure}

\begin{figure}[htb]
\begin{centering}
\includegraphics[width=1.0\columnwidth]{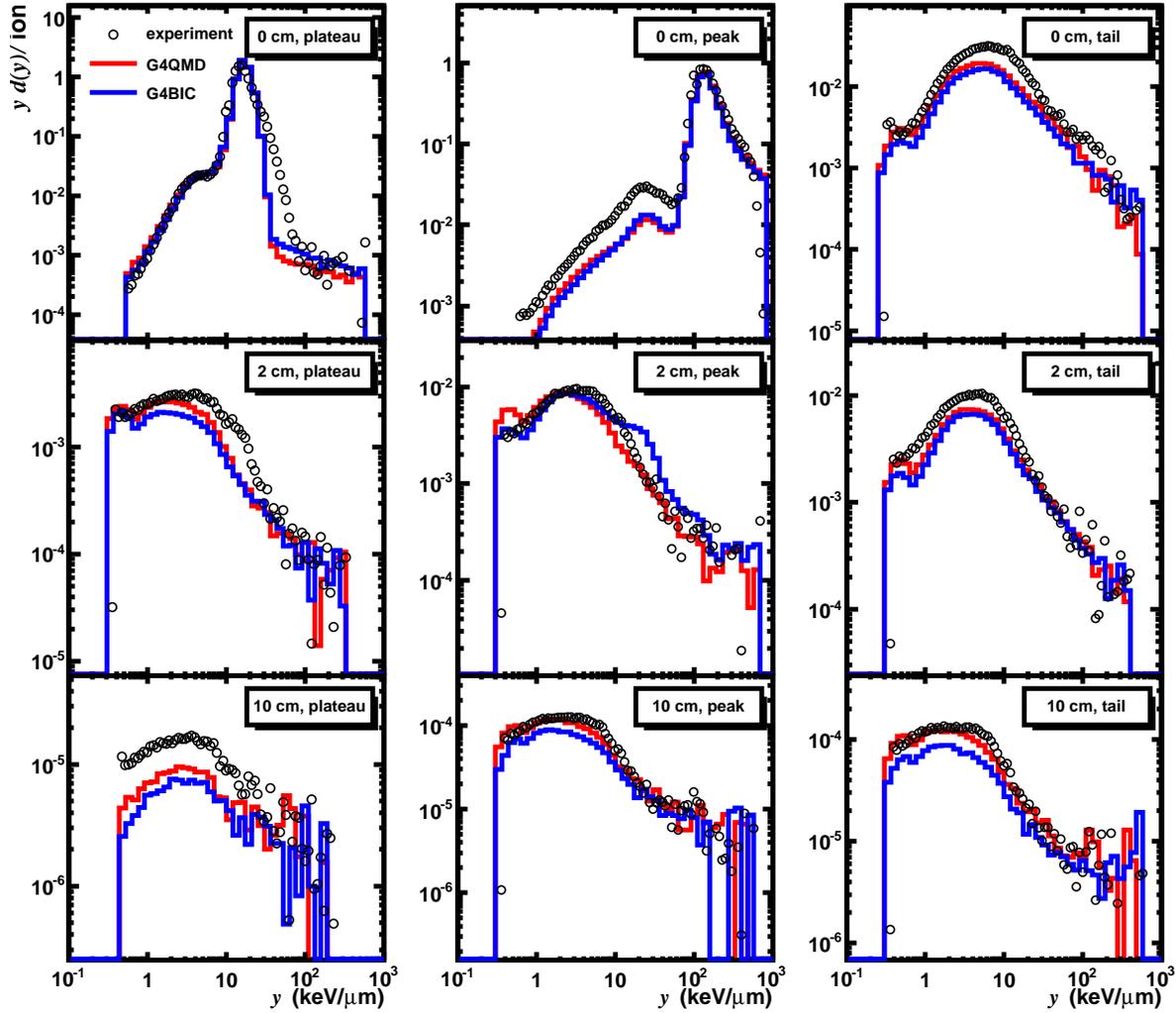}
\end{centering}
\caption{Microdosimetry spectra in water phantom irradiated by
300$A$~MeV $^{12}$C  nuclei calculated by MCHIT with 
standard electromagnetic models including production and transport of $\delta$-electrons
and G4QMD and G4BIC options for nuclear fragmentation,
shown as red- and blue-line histograms, respectively. Points represent experimental
data~\protect\cite{Martino2010}.}
\label{fig:spectra_BICxQMD}
\end{figure}

Following the consideration given above to the particle track patterns in TEPCs, 
in figures~\ref{fig:with_wo_delta_el} and~\ref{fig:spectra_BICxQMD} we present 
calculated microdosimetric $yd(y)$ distributions. 
The experimental data~\cite{Martino2010} were reported only above 0.3--0.5~keV/$\mu$m.
This range was also used to plot calculated microdosimetry spectra and calculate $\bar{y}_f$, 
$\bar{y_d}$ in order to facilitate the comparison with the experimental data.
As discussed in section~\ref{sec:topology}, for the TEPCs placed on the beam axis, 
$\bar{l}>2/3 d$ since a certain number of events is characterized by more than one track traversing
the TEPC, and events with small impact parameter are more frequent.
However, in GSI measurements~\cite{Martino2010} $\bar{l} = 2/3 d$ was assumed
following the standard microdosimetry technique~\cite{ICRU1983}. The same normalization was taken 
in our calculations of $y$ for the sake of a direct comparison with GSI data. 

The MCHIT calculations were performed with different energy threshold for production of delta-electrons
(with no production of delta-electrons taken as a limit case), and also with
two different nuclear fragmentation models.
The sensitivity of calculated distributions 
to computational parameters can be studied by comparing these distributions with each 
other and with experimental data.

The microdosimetric $yd(y)$ distributions measured at GSI~\cite{Martino2010} inside the 
water phantom irradiated by a 300$A$~MeV $^{12}$C beam are shown in 
figure~\ref{fig:with_wo_delta_el} together with our simulation results. 
The distributions are given per beam particle for the same TEPC positions inside and 
outside the beam spot as listed in table~\ref{tab:TEPCpositions} 
and described in section~\ref{sec:topology}.

The influence of the energy threshold for production and transport of $\delta$-electrons 
can be estimated by considering $yd(y)$ distributions
obtained with and without simulating electron production, but with the same nuclear fragmentation 
model (G4QMD), as shown in figure~\ref{fig:with_wo_delta_el}.
Here, the distributions calculated with these two options noticeably differ at 
the ``0~cm, plateau'' TEPC position close to the beam entrance to the phantom. 
In particular, the number of energy deposition events with $20 < y < 200$~keV/$\mu$m is 
much higher compared to the calculation with $\delta$-electrons. At this point 
fast beam nuclei produce energetic electrons which may escape from the TEPC sensitive 
volume and thus reduce the energy deposited to this volume. However, the $yd(y)$ distribution 
calculated at the Bragg peak (at ``0~cm, peak'') is not affected by neglecting $\delta$-electrons. 
Much more secondary electrons are produced close to the Bragg peak, but they are generally 
less energetic compared to those at ``0~cm, plateau'' and do not escape from the TEPC.  
This means that the energy deposited to the TEPC at ``0~cm, peak'' can be simply calculated from 
the stopping power of beam nuclei even without modeling $\delta$-electrons.    
As seen from figure~\ref{fig:with_wo_delta_el}, 
the production of $\delta$-electrons also changes the
distributions in the range of $10 < y < 100$~keV/$\mu$m for the TEPCs located at 2~cm from 
the beam axis.  Similar, but smaller changes in $yd(y)$ distributions associated with electron
production are also found at 10~cm radii. At all these positions energy is deposited to the TEPC 
mostly by protons, and their contribution is reduced if secondary electrons are produced and escape the 
TEPC volume.
The $yd(y)$ distributions with production and transport of $\delta$-electrons are not sensitive
to a change in the lowest energy threshold from 990~eV to 100~eV with Penelope models.

MCHIT results obtained with G4BIC and G4QMD models used to simulate nuclear fragmentation are 
presented in figure~\ref{fig:spectra_BICxQMD} together with experimental data~\cite{Martino2010}.
One has to keep in mind the range of magnitudes of the distributions
presented in nine panels of figures~\ref{fig:with_wo_delta_el} and \ref{fig:spectra_BICxQMD}. 
They are normalized per single beam particle and have a span of almost six orders of magnitude.
This is because of the fact that TEPC hits at the position ``10~cm, plateau'' are approximately
$10^4$ times less frequent than at the Bragg peak (``0~cm, peak''), as described 
below in section~\ref{y_f_section}. The TEPCs located at 10~cm distance
from the beam axis are hit exclusively by secondary particles (mostly nucleons) produced in 
nuclear fragmentation reactions. Therefore, the microdosimetry data~\cite{Martino2010} provide 
another possibility to validate the nuclear fragmentation models of Geant4.

Detailed consideration of the contributions of different
charged particles to the calculated microdosimetric spectra will be given elsewhere~\cite{Burigo}.
In this work only general features of $yd(y)$ distributions are discussed, and
the attention is focused on their general shape due to primary nuclei or secondary
particles. 

In the measurements~\cite{Martino2010} a sharp peak in the spectrum is observed at the TEPC position
``0~cm,~plateau'', figure~\ref{fig:spectra_BICxQMD}.
It is located at $y\sim16$~keV/$\mu$m, and its position is
well reproduced by the MCHIT simulations with both G4BIC and G4QMD. However, as seen from the
corresponding panel of figure~\ref{fig:spectra_BICxQMD}, MCHIT predicts
a slightly sharper drop of $yd(y)$ at the right slope of the peak.
This peak is due to energetic beam nuclei that traverse the TEPC at the entrance to the phantom.
As discussed in section~\ref{sec:topology}, because of a relatively small beam 
diameter (3~mm FWHM) compared to the TEPC size, carbon nuclei propagate close to the 
diameter of the TEPC.
As seen in the panel ``0~cm,~peak'', the peak in the $yd(y)$ distribution corresponding to
$^{12}$C nuclei becomes higher and broader as the TEPC is moved to the stopping 
point of beam nuclei. Compared to the ``0~cm, plateau'' panel, here the peak is shifted
to a much higher value of $y\sim131$~keV/$\mu$m, and it has a prominent satellite peak with
a maximum at $y\sim25$~keV/$\mu$m.  The positions of both peaks are well
reproduced by simulations. The observed shift of the main peak to larger $y$ is due to
increasing the LET of beam nuclei as they are slowing down with their penetration in water.

A common feature of the ``0~cm,~plateau'' and ``0~cm,~peak'' distributions consists in
the presence of additional peaks at lower $y$ values compared to the peaks of beam nuclei.
Such satellite peaks are present due to relatively heavy projectile fragments
(boron, beryllium and lithium nuclei) propagating with velocities
close to the velocity of beam nuclei. Since their $Z^2$ are smaller, their ionization energy 
loss is reduced accordingly. This leads to reduced LET of secondary fragments compared to 
beam nuclei. While the satellite peak is accurately reproduced by
theory at the entrance to the phantom at ``0~cm,~plateau'', a similar peak
at ``0~cm,~peak'' TEPC position is underestimated. This indicates, that the yields of secondary
fragments are underestimated by the nuclear fragmentation models used in MCHIT.

The spectrum at the beam axis beyond the Bragg peak (``0~cm,~tail'') and all the six spectra
at 2 and 10~cm radii do not demonstrate any sharp peaks. This is because of contributions from
various secondary light particles characterized by a broad range of kinetic energies and charges. 
It is expected that recoil charged particles produced in neutron-induced reactions
also contribute to these spectra. This contribution is estimated below in section~\ref{sec:neutrons}.
While the general trends of the  distributions measured at these seven points are 
reproduced by MCHIT, the absolute values are underestimated by a factor of
two or slightly less. In view of a $10^4$ difference in magnitude between ``0~cm,~tail'' and
``10~cm,~plateau'' distributions, such level of agreement between MCHIT and data can be accepted.
Nevertheless, this discrepancy suggests that there is still a room for improvements of
the considered nucleus-nucleus collision models of Geant4, G4BIC and G4QMD, or de-excitation
models.

\subsection{Hit probability, $\bar{y}_f$, $\bar{y}_d$ and dose inside the 
water phantom}\label{y_f_section}

\begin{figure}[htb]
\begin{centering}
\includegraphics[width=0.8\columnwidth]{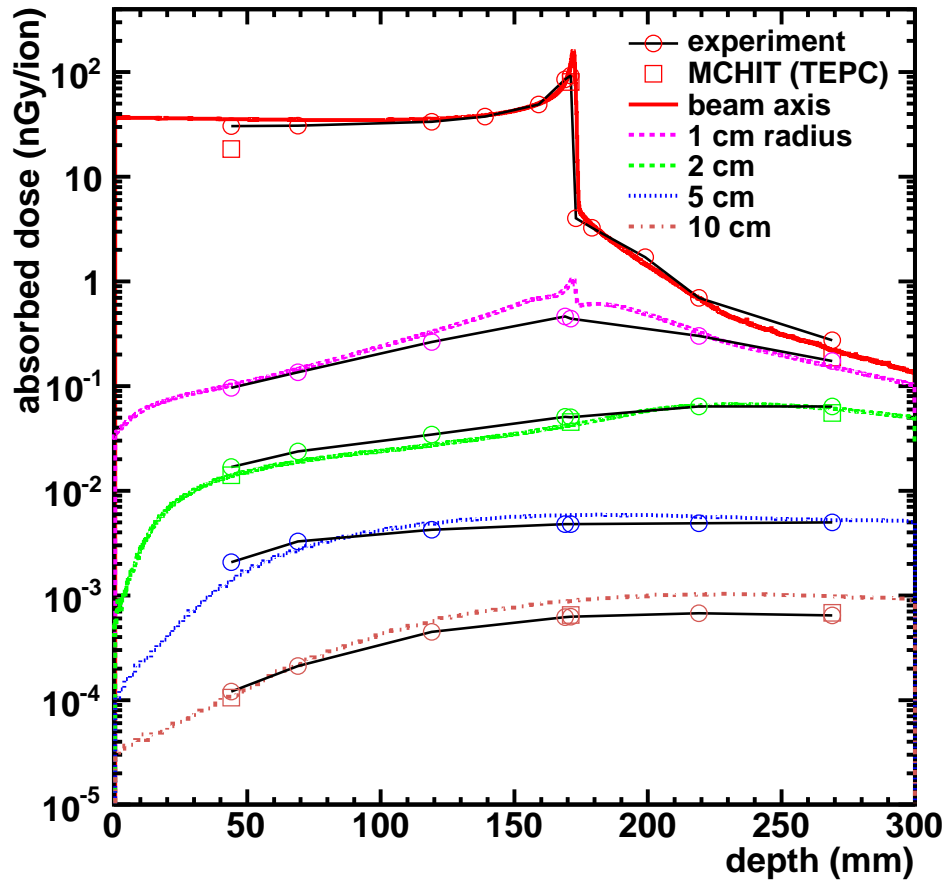}
\end{centering}
\caption{Calculated and measured total doses at various depths in the water phantom 
irradiated by 300$A$~MeV $^{12}$C nuclei. The doses calculated at 0, 2 and 10~cm radii 
from microdosimetry spectra by means of equation~(\ref{eq:dose_practical}) are shown by open squares. 
Lines of various styles explained on the legend present total dose calculated in water by 
direct scoring of energy deposition at 0, 1, 2, 5 and 10~cm radii.   
The measured doses~\cite{Martino2010} are shown by open circles connected by dark solid lines to
guide the eye.}
\label{fig:dose_in_Gy}
\end{figure}

We consider now the average microdosimetric quantities $\bar{y}_f$, $\bar{y}_d$ defined in 
section~\ref{sec:basicsMicro} as well as the total dose at various positions inside the water phantom irradiated  
by 300$A$~MeV $^{12}$C nuclei. The total dose calculated at various distances from the beam 
axis is shown in figure~\ref{fig:dose_in_Gy} as a function of the depth in water together
with measured dose values~\cite{Martino2010}.  

As explained by the legend of figure~\ref{fig:dose_in_Gy}, calculated doses were obtained by two
different methods. In the first method labeled as ``MCHIT (TEPC)'' the dose delivered to the 
tissue-equivalent gas was computed from microdosimetry spectra by calculating firstly $\bar{y}_f$ and 
then dose according to equation~(\ref{eq:dose_practical}).
In the second method calculations were performed in water without placing TEPC.
In the latter method a set of concentric rings was defined inside the phantom 
with 0.1~mm steps in the depth and radius. At the end of each run the energy imparted to each 
ring was calculated and then divided by the mass of the ring and by the 
number of beam particles. Due to a finite size of the TEPC, particles traverse it at various distances
from the beam axis. Therefore, in calculating the dose without TEPC one has to take  
an average dose value for a characteristic volume of a similar size.   
For example, the average dose for 0--6.35~mm radii was calculated to obtain the dose on the beam axis.
The radial ring thickness for dose calculations was set to 2/3 of the TEPC diameter 
at other distances from the beam axis.

The calculated dose values obtained by MCHIT by explicit modeling of the TEPC agree well
with the experimental data at most of the TEPC positions, see figure~\ref{fig:dose_in_Gy}.
However, the dose at ``0~cm, plateau'' is underestimated, as one could expect from
microdosimetric spectrum calculated at this TEPC position where events in the range
$25 < y < 80$~keV/$\mu$m are underestimated, see figure~\ref{fig:with_wo_delta_el}.
The dose values calculated in the water phantom without TEPC agree 
well with the TEPC-based measurements and TEPC-based simulations except 
the points in the vicinity of the Bragg peak. As expected, these points are characterized 
by high dose gradients which are predicted by calculations without TEPC, e.g. at 1~cm radius.
These dose gradients are smoothed by taking average values in water over the above-described
rings. However, the dose in TEPC-based measurements still differs from the 
dose calculated without TEPC in the Bragg peak region at 1~cm radius.

\begin{figure}[htb]
\begin{centering}
\includegraphics[width=0.8\columnwidth]{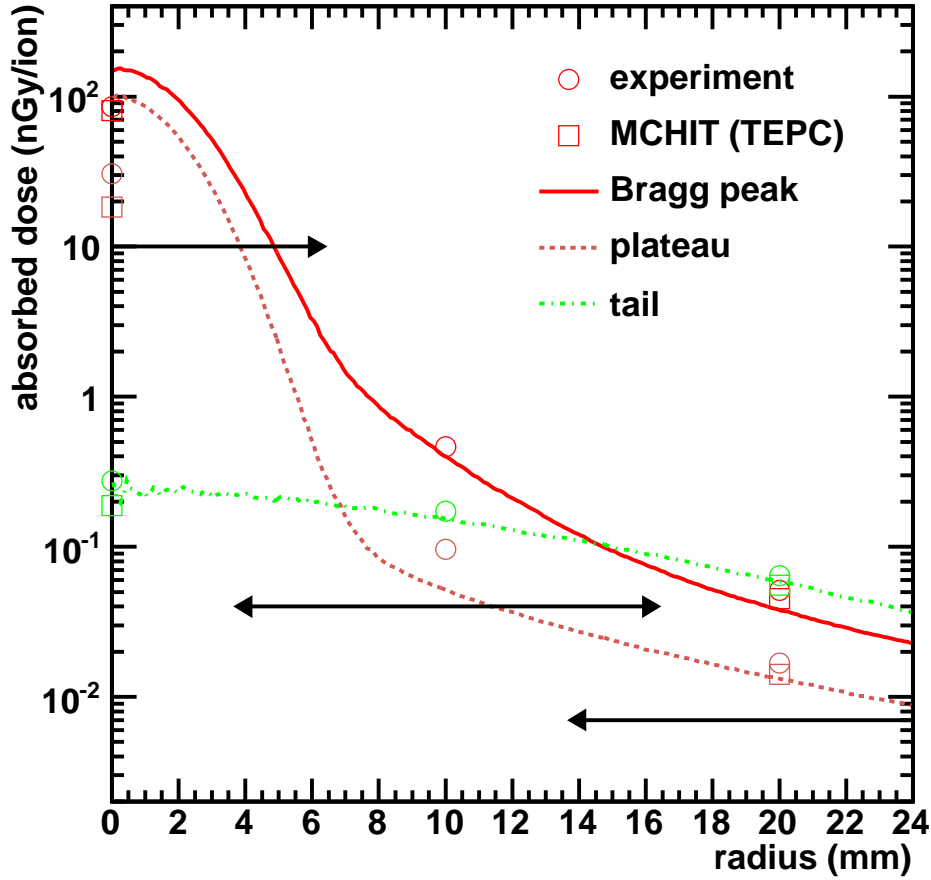}
\end{centering}
\caption{Calculated and measured total dose in the water phantom irradiated by 300$A$~MeV $^{12}$C nuclei
as a function of the distance from the beam axis. The doses calculated on the beam axis
and 20~mm away from it according to equation~(\ref{eq:dose_practical}) are shown by open squares. 
Lines of various styles explained on the legend present total dose calculated in water by 
direct scoring of energy deposition at the plateau (44.04~mm depth), Bragg peak (171.04~mm) 
and tail (269.04~mm). The measured doses~\cite{Martino2010} are shown by open circles.
Lines with arrows demonstrate the radial ranges covered by the TEPC placed at 0, 10 and 20~mm 
from the beam axis.}
\label{fig:dose_in_Gy_radial}
\end{figure}

Figure~\ref{fig:dose_in_Gy_radial} gives further insight into the influence of a finite size 
of TEPC on microdosimetry measurements with therapeutic pencil-like beams. 
The doses from TEPC-based simulations, calculations without TEPC and measured doses 
agree well with each other at 1 and 2~cm distance from the beam axis. The same is true for 
the measurements performed at the beam axis in the tail. In all these cases the spatial
dose distribution is characterized by relatively small gradients inside TEPC volumes which
suggests good agreement between calculations with and without TEPC. 
While a good agreement is observed between data and TEPC-based simulations, direct calculations of the
dose differ from both of them on the beam axis at the plateau and peak. 
As seen, the local dose at the TEPC center is generally 2--3 times higher than the TEPC-measured and 
TEPC-simulated dose under the focused irradiation of this TEPC by 3~mm FWHM pencil-like beam.
Our results suggest that a good agreement between measured and calculated doses can be obtained
only by direct modeling of the TEPC geometry, and not by scoring the dose in simulations without 
TEPC, as done by other authors~\cite{Hultqvist2010,Taleei2011}.

The calculated frequency-mean $\bar{y}_f$ and dose-mean $\bar{y}_d$ lineal
energies are given in tables~\ref{tab:y_f_average} and~\ref{tab:y_d_average}. 
The corresponding values calculated from experimental microdosimetric 
distributions are also given in these tables for comparison.
In addition, the probability $P_{\rm TEPC}$ of energy deposition at a given TEPC position is
also listed in table~\ref{tab:y_f_average}. From these probability values one can conclude, for example,
that on average there is one deposition event per $\sim 3000$ beam particles for 
the TEPCs placed 10~cm away from the beam axis, both at the peak and tail.
The MCHIT results for $\bar{y}_f$ obtained with the Penelope model are higher than those
with the standard electromagnetic model by $1-10$\% depending on the TEPC position.
The measured $\bar{y}_f$ value on the beam axis at the plateau 
is well described by calculations while it is overestimated by $\sim25$\% at the
peak. While calculated with Penelope model $\bar{y}_f$ agree very well with $\bar{y}_f$
measured  at ``2 cm, tail'' and ``10 cm, plateau'' positions, the calculated $\bar{y}_f$ are 
smaller compared to measured values by $10-20$\% at other four positions 
where TEPCs are mostly hit by secondary protons and neutrons.
\begin{table}[htb]
\caption{Probabilities to deposit energy in the TEPC per beam particle at various positions
inside the water phantom and the corresponding $\bar{y}_f$ calculated per energy deposition event.
The calculations were performed   
with Standard Electromagnetic Physics (Option 3) physics list. The results 
obtained with {\em G4EmPenelope} model for transporting electrons, positrons and gammas
are given in the fourth column.
The values of $\bar{y}_f$ calculated from experimental~\cite{Martino2010} 
distributions $yd(y)$ are given for comparison.}
\begin{indented}
\item[]\begin{tabular}{@{}lllll}
\br
  TEPC's        & $P_{\rm TEPC}$    & $\bar{y}_f$     & $\bar{y}_f$     & $\bar{y}_f$     \\
  position      &                   & (keV/$\mu$m)    & (keV/$\mu$m)    &(keV/$\mu$m)     \\
                &   MCHIT           &  MCHIT          & MCHIT           & GSI experiment  \\
                &   Standard opt3   &  Standard opt3  & Penelope        &                 \\
\mr
  0 cm, plateau &$9.87\cdot 10^{-1}$& $14.8$          & $15.2$          & $15.1$          \\
  0 cm, peak    &$6.44\cdot 10^{-1}$& $97.4$          & $99.2$          & $78.7$          \\
  0 cm, tail    &$5.10\cdot 10^{-2}$&\lineup{\0}$2.89$&\lineup{\0}$3.27$&\lineup{\0}$3.52$\\
  2 cm, plateau &$8.25\cdot 10^{-3}$&\lineup{\0}$1.24$&\lineup{\0}$1.34$&\lineup{\0}$1.69$\\
  2 cm, peak    &$2.43\cdot 10^{-2}$&\lineup{\0}$1.39$&\lineup{\0}$1.54$&\lineup{\0}$2.01$\\
  2 cm, tail    &$2.01\cdot 10^{-2}$&\lineup{\0}$2.13$&\lineup{\0}$2.36$&\lineup{\0}$2.41$\\
  10 cm, plateau&$3.19\cdot 10^{-5}$&\lineup{\0}$2.21$&\lineup{\0}$2.18$&\lineup{\0}$2.13$\\
  10 cm, peak   &$3.62\cdot 10^{-4}$&\lineup{\0}$1.33$&\lineup{\0}$1.48$&\lineup{\0}$1.65$\\
  10 cm, tail   &$3.87\cdot 10^{-4}$&\lineup{\0}$1.31$&\lineup{\0}$1.40$&\lineup{\0}$1.59$\\
\br
\end{tabular}
\end{indented}
\label{tab:y_f_average}
\end{table}

The calculated and measured $\bar{y}_d$ values are presented in table.~\ref{tab:y_d_average}. 
As seen from the table,  $\bar{y}_d$ calculated with the standard electromagnetic and Penelope models 
agree well with each other and with experimental data at the beam axis. The difference between calculated
and measured values increases as the distance to the beam axis increases.
As only protons and neutrons contribute for TEPC positions far from the beam, 
this discrepancy points at the necessity to improve the description of production of these secondary
particles by MCHIT.

\begin{table}[htb]
\caption{Calculated and measured~\cite{Martino2010} $\bar{y}_d$ 
per energy deposition event in each TEPC. The calculations were performed   
with Standard Electromagnetic Physics (Option 3) physics list. The results 
obtained with {\em G4EmPenelope} model for transporting electrons, positrons and gammas
are given in the third column.
The experimental values were obtained by integrating $yd(y)$ distributions 
normalized per event.}
\begin{indented}
\item[]\begin{tabular}{@{}llll}
\br
  TEPC's position  & $\bar{y}_d$       & $\bar{y}_d$        & $\bar{y}_d$        \\
                   & (keV/$\mu$m)      & (keV/$\mu$m)       & (keV/$\mu$m)       \\
                   & MCHIT             &  MCHIT             & GSI experiment     \\
                   & Standard opt 3    &  Penelope          &                    \\
\mr
   0 cm, plateau   &\lineup{\0}$16.9$  & \lineup{\0}$17.2$  & \lineup{\0}$18.1$  \\
   0 cm, peak      & $177.$            & $181.$             & $170.$             \\
   0 cm, tail      &\lineup{\0}$13.3$  & \lineup{\0}$13.7$  & \lineup{\0}$14.3$  \\
   2 cm, plateau   &\lineup{\0\0}$6.67$& \lineup{\0\0}$5.68$& \lineup{\0\0}$7.40$\\
   2 cm, peak      &\lineup{\0\0}$8.22$& \lineup{\0\0}$8.81$& \lineup{\0\0}$9.06$\\
   2 cm, tail      &\lineup{\0}$10.4$  & \lineup{\0}$10.6$  & \lineup{\0\0}$9.79$\\
  10 cm, plateau   &\lineup{\0}$14.7$  & \lineup{\0}$12.5$  & \lineup{\0}$13.5$  \\
  10 cm, peak      &\lineup{\0}$11.5$  & \lineup{\0}$16.8$  & \lineup{\0}$10.6$  \\
  10 cm, tail      &\lineup{\0}$11.3$  & \lineup{\0}$15.0$  & \lineup{\0\0}$9.25$\\
\br
\end{tabular}
\end{indented}
\label{tab:y_d_average}
\end{table}

\subsection{Relations between $\bar{y}_f$ and LET for various beam profiles and surrounding 
media}\label{sec:relations_yf_LET}

As discussed in section~\ref{sec:topology}, the TEPC response to focused and homogeneous
irradiation is different. It can be also influenced by the media which surrounds the TEPC, 
as particles produced in this media, e.g. in fragmentation reactions, also hit the detector. 
It is expected~\cite{ICRU1983} that the relation:
\begin{equation}
\bar{y}_f = L \ ,
\label{yf_vs_LET}
\end{equation}
should hold for a spherical TEPC randomly traversed by particles with a constant LET $L$, in the case when 
the production of $\delta$-electrons and other secondary particles is neglected. The relations
between measured, simulated $\bar{y}_f$ and LET under various conditions can be 
assessed by considering table~\ref{tab:y_f_and_LET}. 
In addition to $\bar{y}_f$ calculated
with default physics settings of MCHIT and labeled with (a); the results obtained 
with the Penelope model used for transporting  electrons (b); 
without simulating the production and transport of 
$\delta$-electrons (c); and without simulating nuclear reactions (d) are presented. 

In table~\ref{tab:y_f_and_LET} the results for the TEPC placed
at ``0~cm, plateau'' at the depth of 52.1~mm in water under the impact of 3~mm FWHM beam
at the entrance of the phantom are presented, cases (1a), (1b), (1c) and (1d).
At this TEPC position the energy of $^{12}$C at the entrance to the TEPC volume is
estimated as $\sim218A$~MeV, and it is close to the beam energy ($220A$~MeV) used by 
Taddei et~al. for homogeneous irradiation of a similar TEPC, but surrounded by air~\cite{Taddei2008}.
The thickness of the TEPC wall was twice as large (2.54~mm) compared to the TEPC used at GSI,
and their TEPC operated at different tissue-equivalent gas pressure (effective diameter of $3~\mu$m).
Furthermore, those energy deposition events in the TEPC due to outcoming nuclei different
from the incoming beam nuclei were ruled out in the experimental method. All these details
were taken into account in our simulations.
The comparison of these various irradiation conditions is presented in table~\ref{tab:y_f_and_LET}.

The $\bar{y}_f$ of 14.8~keV/$\mu$m calculated with MCHIT in the case (1a) for TEPC in water is quite  
close to the experimental result of GSI (15.1~keV/$\mu$m) and LET (15.23~keV/$\mu$m). 
In the case (1b) the Penelope model with the production and transport of 
$\delta$-electrons extended to lower electron energy provides a larger $\bar{y}_f$, of 15.2~keV/$\mu$m,
due to the enhancement of energy deposition inside the gas cavity by low-energy $\delta$-electrons
produced in the plastic shell.
In the case (1c) the production of $\delta$-electrons is neglected in the
calculation, and $\bar{y}_f$ increases to 16.2~keV/$\mu$m and exceeds the corresponding LET. 
This is because of additional energy 
deposition in the TEPC sensitive volume which otherwise would be taken away by secondary electrons
propagating beyond the TEPC. The case (1d) presents calculations performed with taking into account
secondary electrons, but neglecting beam fragmentation. This means that instead of the mixture of 
$^{12}$C beam nuclei and secondary fragments, which hit the TEPC in the cases (1a), (1b) and (1c), 
the detector is traversed only by $^{12}$C nuclei and due to their higher $Z^2$ factor,
$\bar{y}_f$ increases to 16.5~keV/$\mu$m, well above the measured $\bar{y}_f$ and also LET.
\begin{table}[htb]
\caption{Calculated and measured  
frequency-mean lineal energy $\bar{y}_f$ per energy deposition event in the TEPC.
Results are given for a Gaussian-shape 3~mm FWHM beam profile (1,2) and for homogeneous irradiation of the
TEPC (3) surrounded by water (1) or air (2,3). Calculations with various 
physics settings are marked by letters: (a) - default MCHIT physics, (b) - the Penelope model 
for electrons, (c) - without simulation of $\delta$-electrons, (d) - without 
simulation of nuclear fragmentation.
Experimental data~\cite{Martino2010}, (1) and  \cite{Taddei2008}, (3) are given for 
comparison with MCHIT results. 
LET values calculated~\cite{Toftegaard2012,Luhr2012} 
according to ICRU Report 73~\cite{ICRU2005} are also given.}
\begin{indented}
\item[]
\begin{tabular}{@{}llllllll}
\br
 \  case & beam   & beam      & media&\multicolumn{3}{c}{$\bar{y}_f$ (keV/$\mu$m) }   & LET \\
         & energy & profile   &      &\crule{3}                                       & (keV/$\mu$m) \\
         &        &           &      &\multicolumn{2}{c}{Calculation} & Experiment    &              \\
         & (MeV/  &           &      &\crule{2}                       &Martino et al. &              \\
         & nucleon)&           &      &   MCHIT       & Taddei et al. & Taddei et al. & ICRU 73      \\
\mr
    (1a) &   218  & Gaussian  & water& 14.8          &                &   15.1        &  15.23       \\
    (1b) &   218  & Gaussian  & water& 15.2          &                &   15.1        &  15.23       \\
    (1c) &   218  & Gaussian  & water& 16.2          &                &   15.1        &  15.23       \\
    (1d) &   218  & Gaussian  & water& 16.5          &                &   15.1        &  15.23       \\
         &        &           &      &               &                &               &              \\
    (2a) &   220  & central   &  air & 20.4          &                &               &  15.14       \\
         &        & incidence &      &               &                &               &              \\
         &        &           &      &               &                &               &              \\
  (3a,d) &   220  & flat      &  air & 13.8          &   14.5         &   13.4        &  15.14       \\
  (3b,d) &   220  & flat      &  air & 14.1          &   14.5         &   13.4        &  15.14       \\
\br
\end{tabular}
\end{indented}
\label{tab:y_f_and_LET}
\end{table}
This also explains the value of $\bar{y}_f$ calculated for the conditions of the second experiment, but with 
central incidence of $^{12}$C nuclei on TEPC, case (2a).
There the beam fragmentation outside the TEPC is negligible as it is surrounded by air. Since in this case 
the TEPC is traversed exclusively by $^{12}$C nuclei along the TEPC diameter, this explains 
the largest energy deposition, and hence the largest $\bar{y}_f$ of 20.4~keV/$\mu$m calculated for this case.  

In order to demonstrate the dependence of $\bar{y}_f$ on the beam shape, the case of a flat beam (3) was also considered.
In this case $\bar{y}_f=13.8$~keV/$\mu$m. This is noticeably smaller than $\bar{y}_f$ of the case (2a), 
20.4~keV/$\mu$m, calculated for the central-incidence of $^{12}$C of the same energy. The ratio between $\bar{y}_f$
for the random impact on the TEPC (3) and for the central beam incidence (2) is remarkably close to
the expected value of 2/3.
One can note that the MCHIT result for the flat beam is closer to the experimental value
than the value calculated by Taddei et al. with another version (7.1) of the Geant4 toolkit.
We attribute this improvement to the changes in electromagnetic models done since the release of the 
version 7.1 and to differences in calculational parameters. However, all calculated and measured $\bar{y}_f$ for the 
flat beam are smaller by $\sim 10$\% compared to LET.

Our results demonstrate that depending on the beam profile the measured $\bar{y}_f$ are either $\sim 35$\% 
higher than corresponding LET for central beam incidence, case (2a), or $\sim 10$\% lower 
than LET for homogeneous irradiation, case (3). At the same time, for the case of TEPC 
irradiation by the Gaussian-shape 3~mm FWHM beam typical for scanning-beam therapy facilities, 
which is an intermediate case with respect to (2a) and (3), the correspondence between calculated 
$\bar{y}_f$ and LET is good. As follows from this analysis, the exact modeling of TEPC irradiation 
conditions (the beam profile and surrounding media) is crucial for reproducing experimental data.

\section{Contributions of secondary neutrons to microdosimetric quantities}\label{sec:neutrons}

The assignment of energy-dependent weighting factors is crucial for 
calculating equivalent doses for neutrons, e.g. in radiation protection.
As discussed in several publications, the calculations of dose from secondary neutrons in 
proton~\cite{Jiang2005,Jarlskog2008,Xu2008} and heavy-ion~\cite{Newhauser2011} 
therapy are prone to various uncertainties. 
First, there are common difficulties for all kinds of radiation in estimating the 
risks from low doses~\cite{Kellerer2000}. Second, depending on the kind of tissue 
under irradiation {\em in vivo} and neutron energy, the relative biological efficiency 
(RBE) varies from 2 to 50~\cite{Grahn1992} or even from 7 to 70~\cite{Dennis1987}.
Third, the fraction of neutrons in complex radiation fields surrounding 
therapeutic beams should be also properly evaluated, as discussed in this section.

The calculations presented above in section~\ref{sec:neutron_measurements} demonstrate that fast neutrons
deposit energy to the tissue-like media mostly by recoil protons. Neutrons produced in fragmentation
reactions during particle therapy can travel long
distances before they interact, thus depositing energy all over the patient's body.
This issue was specially investigated with the MCHIT~\cite{Pshenichnov2005} with the
conclusion that the neutron doses in carbon and proton therapy are at the same order,
about 1\% of the total dose.
As demonstrated by recent measurements~\cite{Yonai2010}, the dose from secondary neutrons in
passive carbon-ion therapy is indeed comparable to the neutron dose in proton
therapy. However, it is not yet clear whether neutrons produced 
during proton or carbon treatments essentially elevate the risk of
secondary cancer~\cite{Jiang2005,Jarlskog2008,Xu2008,Newhauser2011}.  
\begin{figure}[htb]
\begin{centering}
\includegraphics[width=1.0\columnwidth]{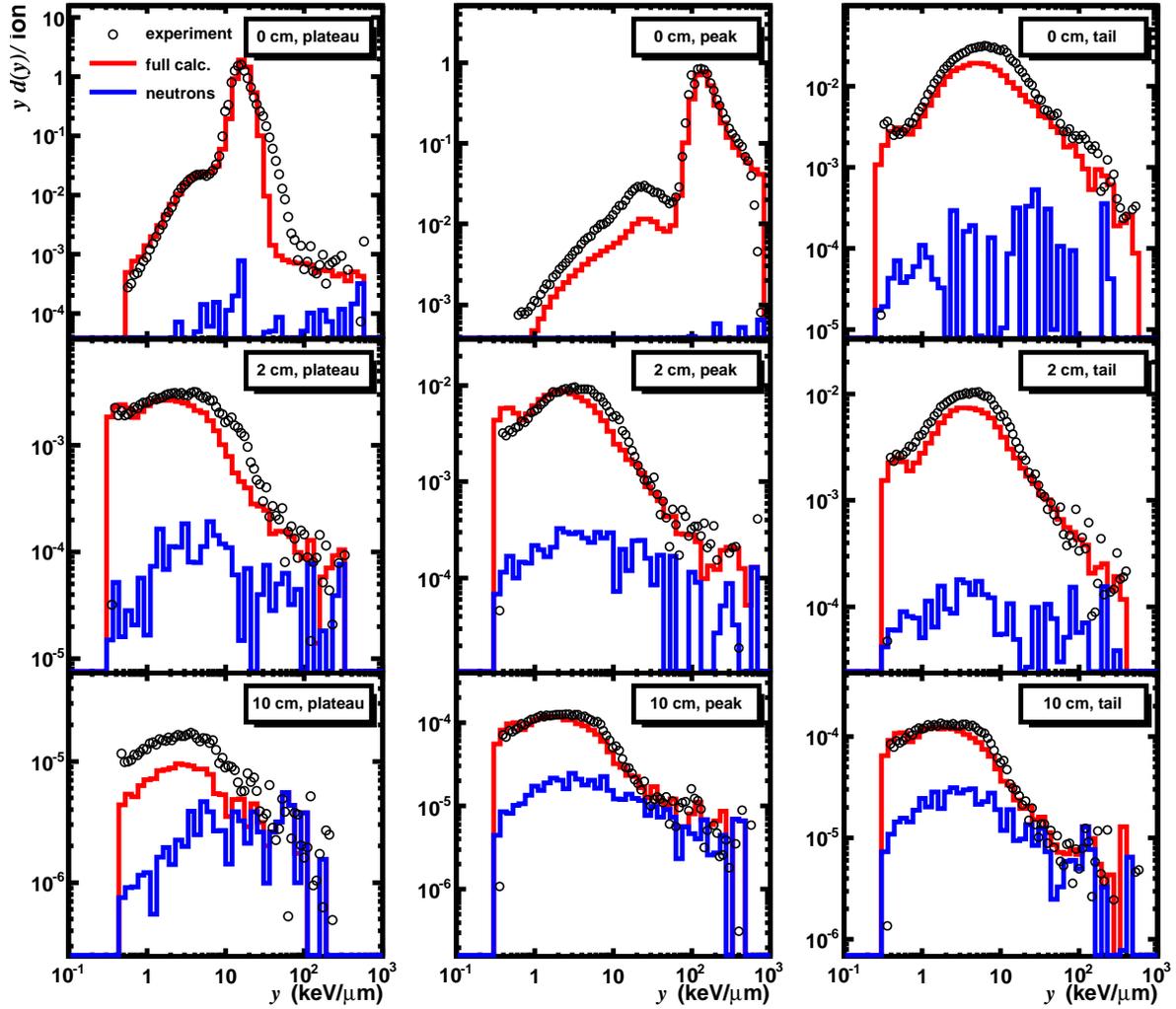}
\end{centering}
\caption{Microdosimetry spectra in water phantom irradiated by
300$A$~MeV $^{12}$C nuclei calculated with MCHIT. Simulations are performed with G4QMD model.
 Points represent experimental
data~\protect\cite{Martino2010}.
}
\label{fig:spectra_neutrons}
\end{figure}
The contributions of secondary neutrons to the microdosimetric spectrum
can be measured applying a veto counter technique which classify events due
to charged or neutral particles. This technique was used elsewhere~\cite{Endo2007a,Wissmann2010}
for 290$A$~MeV carbon beam in acrylic phantom and 200$A$~MeV
carbon beam in water phantom, respectively. Endo et al. found that the neutron
contribution to the deposited dose in the tail was 18\% for the beam axis
and 51\% at 10~cm radius. For larger distances from the beam axis the neutron dose
is predominant, increasing from 74\% in forward direction to 89\% in backward direction at 15~cm radius.

\begin{table}[htb]
\caption{Total dose, dose due to secondary neutrons and the relative neutron contribution to the dose in water phantom 
irradiated by 300$A$~MeV $^{12}$C pencil-like beam.}
\begin{indented}
\item[]\begin{tabular}{@{}lllll}
\br
  TEPC's           & \multicolumn{2}{c}{Total Dose (nGy/ion)}&\multicolumn{2}{c}{Neutrons' Dose (MCHIT)}
                   \\
  position notation& \crule{2} & \crule{2} \\
  & GSI experiment               & MCHIT                       & (nGy/ion)                     & \%               \\
\mr
  0 cm, plateau    & $30.6$                       & $18.4$                       & $<5.0 \cdot 10^{-3}$         &  $<0.03$          \\
  0 cm, peak       & $93.4$                       & $80.6$                       & $<5.0 \cdot 10^{-2}$         & $<0.06$          \\
  0 cm, tail       &\lineup{\0}$2.74\cdot 10^{-1}$&\lineup{\0}$1.88\cdot 10^{-1}$&\lineup{\m}$~1.1\cdot 10^{-3}$& \lineup{\m}$~0.6$\\
  2 cm, plateau    &\lineup{\0}$1.68\cdot 10^{-2}$&\lineup{\0}$1.41\cdot 10^{-2}$&\lineup{\m}$~7.8\cdot 10^{-4}$& \lineup{\m}$~5.5$\\
  2 cm, peak       &\lineup{\0}$5.04\cdot 10^{-2}$&\lineup{\0}$4.52\cdot 10^{-2}$&\lineup{\m}$~2.0\cdot 10^{-3}$&  \lineup{\m}$~4.4$\\
  2 cm, tail       &\lineup{\0}$6.42\cdot 10^{-2}$&\lineup{\0}$5.56\cdot 10^{-2}$&\lineup{\m}$~1.6\cdot 10^{-3}$&  \lineup{\m}$~2.8$\\
  10 cm, plateau   &\lineup{\0}$1.20\cdot 10^{-4}$&\lineup{\0}$1.05\cdot 10^{-4}$&\lineup{\m}$~5.0\cdot 10^{-5}$& \lineup{\m}$47$  \\
  10 cm, peak      &\lineup{\0}$6.25\cdot 10^{-4}$&\lineup{\0}$6.52\cdot 10^{-4}$&\lineup{\m}$~1.5\cdot 10^{-4}$& \lineup{\m}$23$  \\
  10 cm, tail      &\lineup{\0}$6.44\cdot 10^{-4}$&\lineup{\0}$6.81\cdot 10^{-4}$&\lineup{\m}$~1.8\cdot 10^{-4}$& \lineup{\m}$26$  \\
\br
\end{tabular}
\end{indented}
\label{tab:NeutronDose}
\end{table}
In GSI measurements~\cite{Martino2010} contributions to microdosimetric quantities from
secondary neutrons were not identified. Monte Carlo simulations with MCHIT 
make it possible to estimate the energy deposition due to interactions of neutrons. This is performed in two steps.
Firstly, the energy deposition and lineal energy spectra are calculated taking into account 
all relevant physics processes, including production of secondary neutrons and their interactions
with the phantom and TEPC. Secondly, calculations are performed for the same set-up including 
neutron production, but neglecting secondary interactions of neutrons. Thus, the difference between these
two simulations provides the neutron contributions to the dose and $yd(y)$ distributions.  
The absolute dose and relative dose values from neutrons obtained by MCHIT following this procedure are 
given in table~\ref{tab:NeutronDose}. 

The contribution from secondary neutrons
to the microdosimetric spectra is shown in figure~\ref{fig:spectra_neutrons}.
As seen, the relative neutron contribution increases with the distance from the beam axis.
The neutron contribution at 10~cm radius increases from $\sim 25$\% in the forward direction
to $\sim 50$\% in the backward direction. These results can not be directly
compared to those of Endo et al.~\cite{Endo2007a} as they correspond to different TEPC position and 
beam profiles. However, the observed trend of increasing neutron contribution with respect 
to the distance from the primary beam is in agreement with their experimental findings.
One should keep in mind, however, that the absolute doses are very small in these regions.

\section{Conclusions}\label{sec:conclusions}

As stated in a well-known publication on microdosimetry~\cite{ICRU1983}, p 38: ``Experiments and
calculations of microdosimetric spectra play complementary roles. The degree of agreement between
experiment and theory serves as a test of the validity of both.''
Following this approach we extended the MCHIT model to calculations of microdosimetric 
quantities characterizing the radiation effects of accelerated nucleons and nuclei.
The results of the calculations are compared with recent data~\cite{Martino2010} obtained with a therapeutic 
300$A$~MeV $^{12}$C beam and with microdosimetry data collected with quasi-monoenergetic
neutrons~\cite{Nakane2001}. To the best of our knowledge the present study is the first one which uses
the Geant4 toolkit for calculating microdosimetric quantities by a detailed modeling of TEPCs located 
in a phantom both inside and outside of a therapeutic $^{12}$C beam. The in-field and out-of-field
$y$-distributions are obtained in a consistent approach and the contributions of secondary
neutrons to out-of-field doses are evaluated. Our computational methods can be useful to assess the biological
effects of complex radiation fields from therapeutic ion beams, including effects of 
secondary neutrons produced in carbon-ion therapy~\cite{Newhauser2011}. 

Since a TEPC placed in a water phantom irradiated by nuclear beam  inevitably changes 
the amount of material which is traversed by particles, the radiation field is distorted 
in the presence of the detector. The TEPC geometry was thoroughly implemented in the Monte Carlo
simulations performed in this work with the MCHIT model.
Realistic models from Geant4 toolkit were used to describe particles propagation and
energy deposition in non-uniform medium.
This allowed us to obtain 
a good agreement between calculated and measured microdosimetric quantities
for different TEPC positions inside the water phantom irradiated by the pencil-like beam
of 300$A$~MeV $^{12}$C nuclei. Our main conclusions are as follows:

\begin{itemize}

 \item The MCHIT model is able to describe the spatial distribution of the total dose in the water phantom
despite of six orders of magnitude decrease of dose with increasing distance from the beam axis. 

 \item The contributions of delta-electrons on energy deposited to TEPC varies at 
different TEPC locations in the water phantom. The propagation of energetic 
beam nuclei through a TEPC is accompanied by production of energetic delta-electrons, 
which may escape the TEPC sensitive volume, thus reducing the deposited energy, 
while this effect is less important for the detectors impacted only by 
secondary nucleons, which produce low-energy electrons far from the beam 
axis.

 \item Contributions of primary beam nuclei and secondary fragments can be distinguished in 
the calculated and measured $yd(y)$ microdosimetric spectra on the beam axis. 

 \item Proper modeling of nuclear fragmentation reactions is crucial for describing microdosimetric 
distributions both on the beam axis and far from the beam. The nuclear fragmentation models of Geant4,
G4QMD and G4BIC, are equally suitable for describing general features of the spectra, but both 
underestimate the fluxes of protons and neutrons far from the beam. This indicates the necessity of improving
nucleus-nucleus collision models in calculating the angular and energy distributions of 
secondary nucleons and nuclear fragments.    

 \item The values of $\bar{y}_f$ and $\bar{y}_d$ for a TEPC under direct impact of projectile 
nuclei are sensitive to the beam profile.   

 \item The MCHIT model describes well the $yd(y)$ distributions in the PMMA phantom 
irradiated by quasi-monoenergetic neutrons. This justifies the use of this model for calculating the 
TEPC response to secondary neutrons from $^{12}$C beam.

 \item The contribution of secondary neutrons to the out-of-field dose from $^{12}$C beams estimated from 
MCHIT simulations amounts to about 50\% of the total far from the beam.
Since experimental identification of neutrons would require
bulky detectors placed only outside the water phantom, such microdosimetry measurements 
remain the only solution to estimate the upper limits for the dose from neutrons and their 
radiation quality close to the target volume. 

\end{itemize}

Our results can be extended to other measurements with TEPC, in particular with other kinds of 
nuclear beams, as the TEPC model (Far West Technology Inc., LET-1/2) simulated in this work is  
used worldwide. A detailed consideration of contributions of specific nuclear fragments to
$yd(y)$ distributions calculated with MCHIT will be given elsewhere~\cite{Burigo}.

\ack
The presented results were obtained in the framework of NanoBIC-NanoL project.
L.B. is grateful to the Beilstein Institute for support. This work was also partially supported by HIC for FAIR within the Hessian LOEWE-Initiative.
We wish to thank D.~Schardt, G.~Martino, C.~La~Tessa and M.~Durante for numerous discussions 
which inspired us to conduct this theoretical study and also for 
providing us with their tables of experimental data.
Our calculations were performed at the Center for Scientific Computing (CSC) of the
Goethe University Frankfurt. We are grateful to the staff of the Center for support.

\section*{References}


\end{document}